\documentclass{article}[11pt]
\usepackage{jheppub}

\usepackage[toc,page]{appendix}
\usepackage{amsmath,amssymb}
\usepackage{bm}
\usepackage{booktabs}
\usepackage{chemformula}
\usepackage{color}
\usepackage{cleveref}
\usepackage{dcolumn}
\usepackage{enumitem}
\usepackage{epstopdf}
\usepackage{epsfig}
\usepackage{graphicx}
\usepackage{hyperref}
\usepackage{siunitx}
\usepackage{todonotes}
\usepackage{url}
\usepackage{wrapfig}
\usepackage{xcolor}
\ifdefined\qtyproduct
\else
  \ifdefined\NewCommandCopy
    \NewCommandCopy\qtyproduct\SI
  \else
    \NewDocumentCommand\qtyproduct{O{}mm}{\SI[#1]{#2}{#3}}
  \fi
\fi

\usepackage{macros}

\toccontinuoustrue
\begin{document}


\title{Centralised Design and Production of the Ultra-High Vacuum and Laser-Stabilisation Systems for the AION Ultra-Cold Strontium Laboratories}

\abstract{
This paper outlines the centralised design and production of the Ultra-High-Vacuum sidearm and Laser-Stabilisation systems for the AION Ultra-Cold Strontium Laboratories. Commissioning data on the residual gas and steady-state pressures in the sidearm chambers, on magnetic field quality, on laser stabilisation, and on the loading rate for the 3D Magneto-Optical Trap are presented. Streamlining the design and production of the sidearm and laser stabilisation systems enabled the AION Collaboration to build and equip in parallel five state-of-the-art Ultra-Cold Strontium Laboratories within 24 months by leveraging key expertise in the collaboration. This approach could serve as a model for the development and construction of other cold atom experiments, such as atomic clock experiments and neutral atom quantum computing systems, 
    by establishing dedicated design and production units at national laboratories.\\
\begin{flushright}
\vspace{-5mm}AION-REPORT/2023-03
\end{flushright}}

\author[]{{\large AION Collaboration}:\\~~\\}
\author[]{B.~Stray, O.~Ennis, S.~Hedges, S.~Dey, M.~Langlois, K.~Bongs, S.~Lellouch, M.~Holynski;~$^a$\\\vspace{1.5mm}}
\author[]{B.~Bostwick, J.~Chen, Z.~Eyler, V.~Gibson, T.~L.~Harte, M.~Hsu, M.~Karzazi, J.~Mitchell, N.~Mouelle, U.~Schneider, Y.~Tang, K.~Tkalcec, Y.~Zhi;~$^b$\\\vspace{1.5mm}}
\author[]{K.~Clarke and A.~Vick;~$^c$\\ \vspace{1.5mm}}
\author[]{K.~Bridges, J.~Coleman, G.~Elertas, L.~Hawkins, S.~Hindley, K.~Hussain, C.~Metelko, H.~Throssell;~$^d$\\\vspace{1.5mm}}
\author[]{C.~F.~A.~Baynham, O.~Buchm{\"u}ller, D.~Evans, R.~Hobson, L.~Iannizzotto-Venezze, A.~Josset, E.~Pasatembou, B.~E.~Sauer, M.~R.~Tarbutt;~$^e$\\\vspace{1.5mm}}
\author[]{L.~Badurina, A.~Beniwal, D.~Blas,~\footnote{Present address: Grup de F\'isica Te\`orica, Departament de F\'isica, Universitat Aut\`onoma de Barcelona, 08193 Bellaterra (Barcelona), Spain and Institut de Fisica d’Altes Energies (IFAE), The Barcelona Institute of Science and Technology, Campus UAB, 08193 Bellaterra (Barcelona), Spain} J.~Carlton, J.~Ellis, C.~McCabe;~$^f$\\ \vspace{1.5mm}}
\author[]{E.~Bentine, M.~Booth, D.~Bortoletto, C.~Foot, C.~Gomez, T.~Hird, K.~Hughes, A.~James, A.~Lowe, J.~March-Russell, J.~Schelfhout, I.~Shipsey, D.~Weatherill, D.~Wood;~$^g$\\\vspace{1.5mm}}
\author[]{S.~Balashov, M.~G.~Bason, J.~Boehm, M.~Courthold, M.~van~der~Grinten, P.~Majewski, A.~L.~Marchant, D.~Newbold, Z.~Pan, Z.~Tam, T.~Valenzuela, I.~Wilmut~$^h$\\ \vspace{1.5mm}}
\affiliation[a]{Physics and Astronomy, University of Birmingham, Edgbaston, Birmingham, B15 2TT, UK}
\affiliation[b]{Cavendish Laboratory, J J Thomson Avenue, University of Cambridge, CB3 0HE, UK}
\affiliation[c]{ASTeC, STFC Daresbury Laboratory, Warrington, WA4 4AD, UK}
\affiliation[d]{Department of Physics, University of Liverpool, Merseyside, L69 7ZE, UK}
\affiliation[e]{Department of Physics, Blackett Laboratory, Imperial College, Prince Consort Road, London, SW7 2AZ, UK}
\affiliation[f]{Physics Department, King's College London, Strand, London, WC2R 2LS, UK}
\affiliation[g]{Department of Physics, University of Oxford, Keble Road, Oxford, OX1 2JJ, UK}
\affiliation[h]{STFC Rutherford Appleton Laboratory, Didcot, OX11 0QX, UK}

\maketitle
\tableofcontents
\newpage

\section{Introduction}
\label{sec:intro}

The Atom Interferometer Observatory and Network (AION) project~\cite{AION,AIONweb} is developing quantum technologies using cold atoms for a novel class of fundamental physics experiments. These will enable searches for ultralight dark matter and measurements of gravitational waves from astrophysical and cosmological sources in the deci-Hz frequency band that is inaccessible to other current and planned experiments~\cite{Badurina:2021rgt}. These sources may include black holes with masses intermediate between those discovered by the LIGO and Virgo experiments and the heavier black holes known to be present in galactic nuclei~\cite{doi:10.1146/annurev-astro-032620-021835}.

AION will demonstrate the required deployable and scalable quantum technology by constructing and operating staged \SI{10}{\meter}- and \SI{100}{\meter}-scale instruments, paving the way for a future km-scale facility and a possible spaceborne atom interferometer experiment such as AEDGE~\cite{AEDGE}. In the first phase of stage 1 of the AION project, funded via the UKRI QTFP programme~\cite{UKRIQTFP}, a joint STFC and EPSRC team has successfully delivered a new collaborative network in cold-atom interferometry, and by spring 2024 this collaboration will have made key steps towards the construction and deployment of a first 10-m prototype instrument.

The AION project requires the state of the art in single-photon strontium atom interferometry to be developed to new levels. To this end, one of the main deliverables of the AION project in its first phase of funding was to build dedicated Ultra-Cold Strontium Laboratories (UCSLs) tasked with conducting complementary R\&D efforts in parallel, at the University of Birmingham (Large Momentum Transfer), the University of Cambridge (transport, cooling, launch and atom optics), Imperial College London and the Rutherford-Appleton Laboratory (RAL) (squeezing and high-intensity atom interferometry), and the University of Oxford (atom sources and location of the AION-10 detector).

Coordinating the parallel delivery of five state-of-the-art UCSLs within 24 months represented a new challenge for cold atom physics, and is a keystone of any future cold-atom infrastructure project.
In order to accomplish this ambitious goal, the AION project adopted an approach that is well established in, e.g., the particle physics community, of centralising the design and production of key components of the UCSLs, namely the Ultra-High Vacuum (UHV) and Laser-Stabilisation (LS) systems.

Traditionally, such systems are designed and built in-house by each laboratory, requiring significant staff and time resources in each location. By leveraging key expertise within institutions in the collaboration, such as in LS at Imperial College London, in oven and source design at the University of Oxford, in static and tunable-focus dipole traps at the University of Cambridge, and the expertise in UHV construction at the RAL and Daresbury national laboratories, it was possible to streamline the design and production of both UHV and LS systems~\footnote{With the exception of the Imperial UHV system, which required specialised in-vacuum hardware for quantum squeezing measurements.}. This required establishing a high level of collaboration across all UCSLs, which greatly shortened the resource required at each individual institution to build the five UCSLs compared to the traditional approach.

This document summarises the approach and methods utilised for the centralised design and production of the UHV and LS systems for the AION UCSLs~\footnote{The optical designs and control systems were also centralised.}, which could serve as a useful model for the modular and distributed development and construction of other cold atom experiments, such as atomic clock experiments or neutral atom quantum computing systems. Commissioning data on the performances of the UHV and LS systems are also presented. The AION project is currently investigating the possibility of commercialising the centralised design and production of the UHV and LS systems, facilitating broader use of this approach for other cold atom 
projects.

\section{Laser Cooling of Strontium - Requirements}
\label{sec:laser_cooling_Sr}
\begin{figure}[h]
    \centering
    \includegraphics[width = 0.95\textwidth]{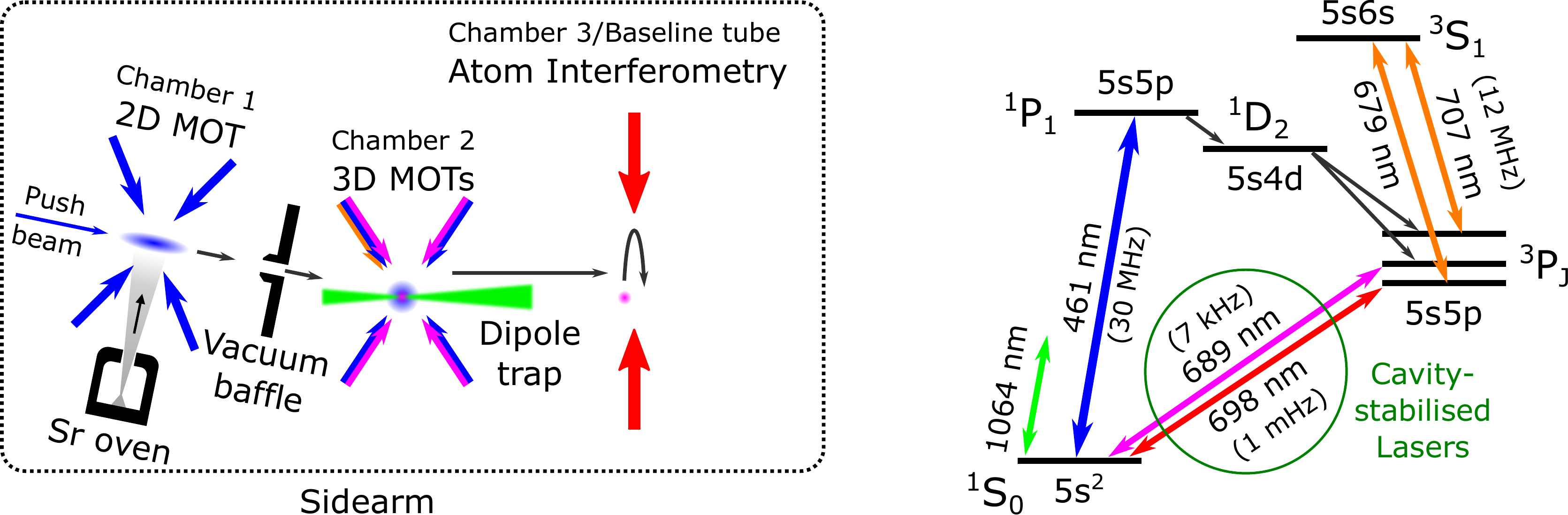}
    \caption{
    \textit{Left:} Conceptual schematic of an AION sidearm, depicting the strontium laser cooling regions. The Magneto-Optical Traps (MOTs) are created inside ultra-high vacuum chambers (see Section~\ref{sec:UHV}), and require several laser beams (see Section~\ref{sec:lasers}) as well as externally-applied magnetic fields (See Section~\ref{sec:magnets}). 
        \textit{Right:} Strontium level scheme, showing the laser wavelengths used for cooling, trapping, and interferometry. Linewidths are given in brackets. The 5s5p~\textsuperscript{3}P\textsubscript{0} state has \SI{1}{\milli\hertz} linewidth  for the \textsuperscript{87}Sr isotope used in AION; in the even isotopes \textsuperscript{88,86,84}Sr the state is much narrower  because single-photon decay is strictly forbidden.}
    \label{fig:laser_cooling_diagrams}
\end{figure}

In the planned atom interferometers the atoms will be cooled and prepared in dedicated UHV chambers mounted to the side of the vertical main interferometer tube before being transported into the main tube for launching and atom interferometry~\cite{AION}. These auxiliary chambers are referred to as \textit{sidearms} and their specifications regarding structure, magnet systems, and parts of the laser stabilisation system are driven by the  
laser cooling techniques used to generate cold strontium clouds. The laser cooling scheme employed by AION is depicted in Fig. \ref{fig:laser_cooling_diagrams}.

The AION cooling sequence follows a similar pattern as previous experiments with cold strontium \cite{stellmer_production_2013,desalvo_degenerate_2010}:

\begin{enumerate}
    \item Effusion of atoms from a hot strontium oven;
    \item Capture of atoms in a 2D ``blue" Magneto-Optical Trap (MOT) in Chamber 1, creating a beam of slow atoms;
    \item Recapture of the slow beam in a 3D ``blue" MOT in Chamber 2;
    \item Further cooling to \SI{1}{\micro\kelvin} in a 3D ``red" MOT;
    \item Loading into an optical dipole trap for evaporative cooling;
    \item Optical transport into the atom interferometry region (either Chamber 3 in test sidearms, or the main vertical tube in the full-scale detector);
    \item Atom launch, interferometry, and measurement.
\end{enumerate}
The layout of the three-chamber vacuum systems at the Universities of Cambridge and Oxford is shown in Fig.~\ref{fig:vacuum_system}.

All stages in the cooling sequence must be carried out in an ultra-high vacuum environment, to reduce background gas collisions to a negligible rate. Pressures of $\lesssim 10^{-10}$~mbar are required in Chambers 2 and 3 to ensure negligible background atom loss during evaporative cooling in Chamber 2 and atom interferometry in Chamber 3. The AION vacuum system meeting these requirements is described in Section~\ref{sec:UHV}.

For stages 1-3 of the sequence, strong magnetic field gradients in Chambers 1 and 2 are required. In Chamber 1, these fields can be applied permanently through the use of NdFeB permanent magnets that reduce power consumption and cooling requirements. In the case of Chamber 2, high-current water-cooled coils are needed to make the required magnetic field
gradient - see Section~\ref{sec:magnets}.

The stages of the sequence addressing narrow transitions (red MOT; atom interferometry) require narrow-linewidth, cavity-stabilised lasers at \SI{689}{\nano\meter} and \SI{698}{\nano\meter}, whereas the stages that address broader transitions (blue MOT; repumping transitions) or use far off-resonance light (dipole trap) rely on wavemeter-locked lasers at \SI{461}{\nano\meter}, \SI{679}{\nano\meter}, \SI{707}{\nano\meter}, and free-running lasers at \SI{1064}{\nano\meter} - see Section~\ref{sec:lasers}.

\section{Sidearm System}
\label{sec:sidearm}
\subsection{Ultra-High Vacuum System} \label{sec:UHV}

\subsubsection{Overview}
\label{sec:overview}
\begin{figure}[t]
    \centering
    \includegraphics[width = 0.9\textwidth]{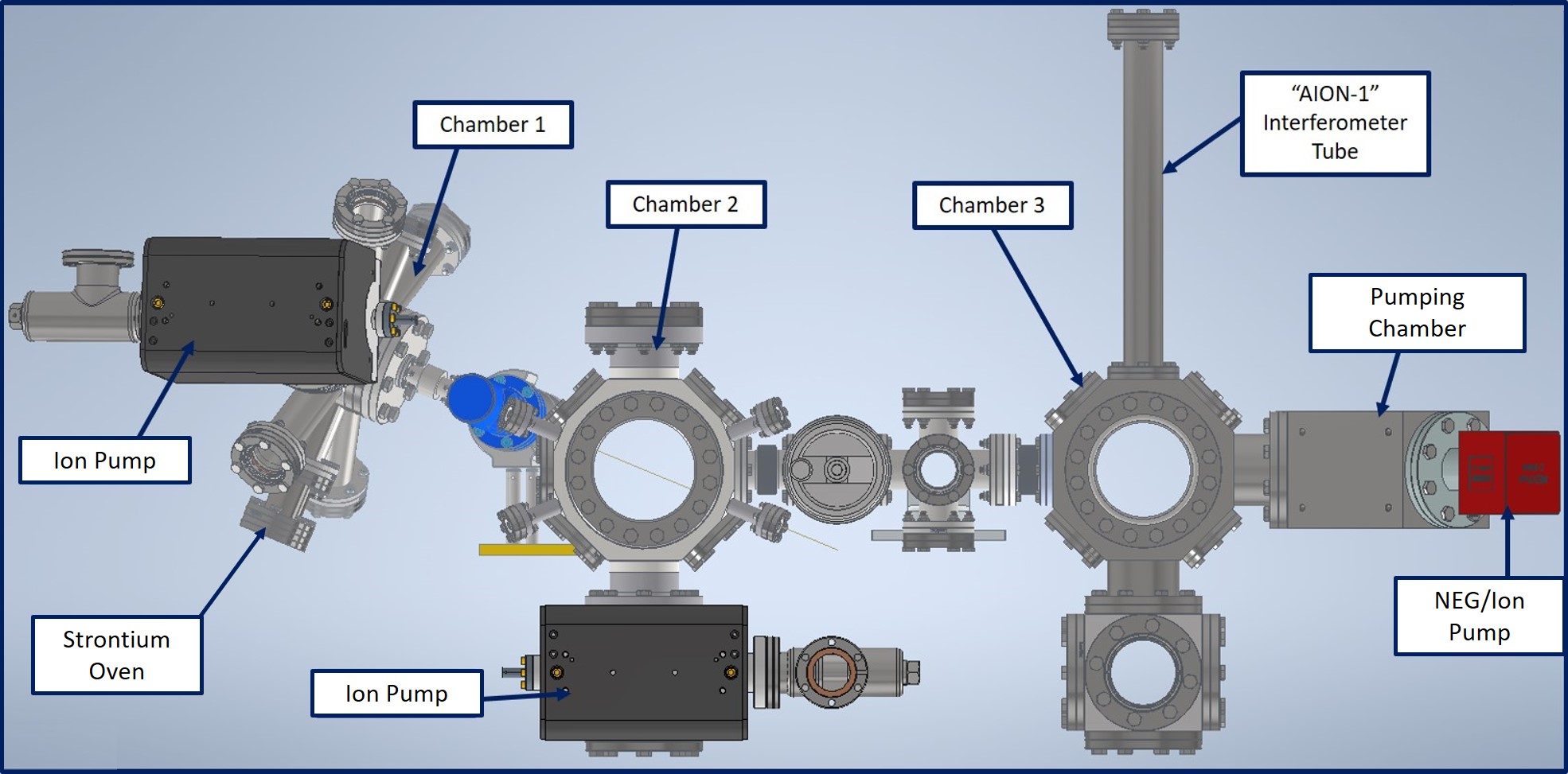}
    \caption{The layout of the three-chamber vacuum systems at the Universities of Cambridge and Oxford.} 
    \label{fig:vacuum_system}
\end{figure}

Although the overall designs of the five vacuum systems are similar,
their precise designs vary slightly according to the intended R\&D work to be undertaken with each sidearm. Unless otherwise mentioned, this document describes the full three-chamber sidearm used at the Universities of Oxford and Cambridge. The sidearms used at Imperial College, RAL and the University of Birmingham do not include Chamber 3, and the Birmingham sidearm includes an extra six-way cross and tube on Chamber 2 to facilitate development of large momentum transfer atom interferometry sequences.

\begin{figure}[t]
    \centering
    \includegraphics[width = 0.35\textwidth]{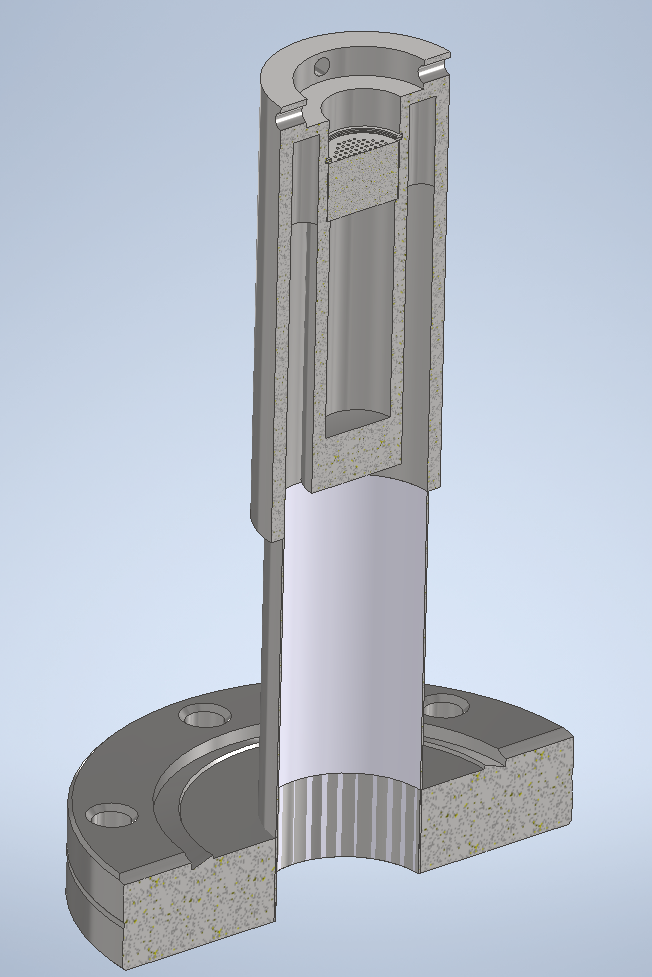}
    \caption{The stainless steel oven for heating strontium. The six cartridge heaters and two thermocouples (not shown) are inserted from the rear (bottom of drawing) without the need for any electrical feedthroughs. A thin-walled stainless steel tube of low thermal conductance connects the cylindrical oven body to a Conflat flange (CF40). The nozzle was laser machined from a stainless steel disc and held in place by a C-clip (150 circular channels of 0.3~mm  diameter and 3~mm long), and mounted within a recess in the oven body to reduce thermal losses (cf. mounting flush with the end of the cylinder).} 
    \label{fig:oven}
\end{figure}

Chamber 1, shown in Fig.~\ref{fig:vacuum_system}, contains the strontium oven, which is illustrated in Fig.~\ref{fig:oven}, whose presence limits its attainable vacuum pressure. For this reason, Chambers 1 and 2 have separate ion pumps with the transfer line between the two containing a differential pumping aperture of \SI{3}{\milli\meter} diameter and \SI{15}{\milli\meter} length. Chamber 1 has multiple viewports, four of which are in the required configuration for the 2D MOT. These are outside the line-of-sight of the oven, so as to ensure that they are not coated with strontium when the oven is operating. The background vacuum level in Chamber 1 was required to be lower than $10^{-9}~$\SI{}{\milli\bar} in order not to affect the base pressure in the adjoining chamber.
The chamber is processed, built and tested to the same standards as the rest of the vacuum system.

Chamber 2 includes the 3D MOT chamber and the one or two cubes that sit below, depending on the system specification. All of Chamber 2 is pumped by a single 20L/s ion pump and a NEG cartridge pump that connects to the cube directly beneath the chamber. Chamber 3 is located beyond Chamber 2, connected through a gate valve, and the transition from Chamber 2 to Chamber 3 represents the transition from the sidearm to the interferometer tube. Chamber 3 (present only in the Oxford and Cambridge assemblies) has no magnets and only optical access. Chamber 3 is used, along with its associated extension tube, as a proxy for a 1~m interferometer.  The launch optics are housed around the 6-way cube below the chamber.  Chamber 3 is pumped from a final `pump chamber' with a NEG-ion pump.

\subsubsection{Design}
\label{sec:design}

The design of the vacuum system and associated support frame was driven by a variety of requirements. These included the common requirements of such a system, such as vacuum performance, ability for bake-out, access to the magnet system, optical access, and operational stability. In addition, the centralised production drove additional requirements, namely strength and stability for transport, installation and servicing access and methodology at the destination institutions and the ability to manufacture efficiently  or procure vacuum components.
The solutions to the above requirements benefited from a centralised engineering approach in a number of ways.

A key metric for the vacuum systems is their vacuum quality, as collisions with background gas atoms ultimately limit the lifetime of ultracold atoms and can thereby limit the duration of the experimental sequence.
Furthermore, it was critical that this performance be consistent across all five systems, as inconsistent vacuum levels between the five systems could lead to  difficulties when transitioning technology between the sidearms.  The ability to transfer technology seamlessly between institutions is a major requirement of the systems, as key deliverables rely on the integration of the technologies developed in the five UCSLs into a single system to demonstrate interferometry on the clock transition.  Centralising the design of the system and specification of the pumping strategy ensured that the project optimised the prospects for achieving the required similarity of performance between systems.

Ancillary system integration had to be considered at a fundamental and conceptual design level, because these systems interface very closely with the vacuum system.  A poorly-considered fastener arrangement or an unstable mount could ultimately cause a catastrophic leak in the system.  This was especially relevant for the permanent magnets and electromagnetic coils required for the MOTs  in Chambers 1 and 2 (see Section~\ref{sec:magnets} for more detailed information on these systems).  These magnet systems are installed after the vacuum system has been sealed, evacuated, baked out and signed off, and any failure of the vacuum system during installation might require transport back to the build facility, partial-rebuild and re-bake.  A particularly striking example of this is the installation of the electromagnetic coils for the Chamber 2 MOT (the 3D MOT).  Installation of one half of these coils presented a potential issue in the form of a clash between the coils and the gate valve separating Chambers 1 and 2.  A common modular mounting system was therefore designed for all sidearms, which would allow safe installation of the coils retrospectively onto the vacuum system.


Subsystems and components that require methodical or detailed installation methods and techniques benefit greatly from the centralised approach as individual laboratories avoid operating at an elevated risk during the installation process.
Standardisation of viewport specification was an important example of this. 
An effort was made to ensure, where possible, that all systems had the same specifications for viewports in the same position.  
This allowed for central procurement and a centralised inventory of viewports that led to a decreased level of risk during the integration phase of the project, which would not have been possible had each laboratory been specifying independently the viewports required for their deliverables.  
The centralised inventory became invaluable during the build process when the fragility and attrition rate of the viewports became apparent during the first bakes:  A substantial risk to the fragile viewports was the bakeout temperature profile, and the centralised, sequential builds enabled us to improve progressively  the bakeout profile for later bakes, thereby substantially reducing the risk associated with the bake-out cycle. See Section~\ref{sec:Processing_Vacuum_System} for more information on the centralised build and its benefits.

\begin{figure}[t]
    \centering
    \includegraphics[width = 0.9\textwidth]{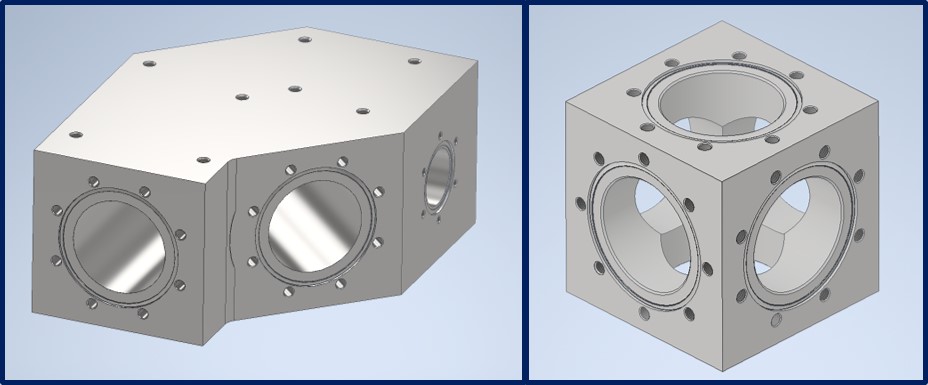}
    \caption{\textit{(Left)} The NEG pump chamber and \textit{(right)} the interferometer cube, which were fabricated from 316L using EDM as described in the text.}
    \label{fig:EDM_Cham}
\end{figure}

During design of the vacuum chambers for the system - see Fig.~\ref{fig:EDM_Cham} 
for images of these chambers - the mechanical engineering team at the University of Oxford looked into the effects of material selection on magnetic permeability.  Traditionally 316LN (re-annealed after machining or fabrication) has always been specified for vacuum systems where low magnetic permeability has been required.  The Oxford team looked into the effects of the inherent stresses of traditional machining (milling, turning, honing, grinding, etc.) on the material and how re-annealing the material relaxes and dissipates these stresses.  It was observed that reducing the stresses experienced by the material during the manufacturing process resulted in a reduced magnetic permeability of the final part. In the end the mechanical engineering team was able to use Electrical Discharge Machining (EDM), a force-free machining process, to create low magnetic permeability vacuum components out of 316L at a reduced cost.  Wire cutting was used for the majority of the machining, using spark erosion with a custom electrode in some places, to form non wire-cut features.  The centralised approach facilitated this quasi-independent research, and the bulk manufacture of components for vacuum systems. Each system benefited performance-wise and economically from this work, making this endeavour cost-neutral.

\subsubsection{Designing the sidearm packaging}
\label{sec:frame}

From the onset of the project, the sidearms were conceived as development platforms towards the AION-10 and AION-100 projects. A sidearm for an AION-10 or AION-100 application would require packaging. The choice to package the sidearm into a single transportable object that could be delivered to each institution (or by extension an AION-10 or AION-100 construction site), was one of the key decisions that enabled the manufacture to be centralised.

The constraints on the packaging of the sidearm were minimal. It needed to be self-contained, transportable, robust, and large enough to accommodate most or all of the optical systems that would need to be integrated around it.

The initial sidearm packaging concept included an optical table as the base on which the vacuum system and other components would be assembled. In order to increase their flexibility as a development platform and to stay closer to the eventual requirements of AION-100, a choice was taken to simplify this and instead to mount the packaged sidearm on existing optical tables.
With hindsight this choice complicated matters considerably, as there needed to be many removable parts, and a complex transfer process to fit the sidearm onto existing optical tables.

There were five sidearms built in a total of four varieties so as to enable the exploration of different areas of the sidearm development in different institutes, as discussed above. The design started with the package for the Universities of Oxford and Cambridge, which was the most complex, and packaged it in such a way that it could be shortened for RAL and Imperial College, and could in turn be made taller for the University of Birmingham. The four versions of the packaging frame are all based on a common design and the variations are well understood, enabling further developments of laser systems and system operation. 
Although the customisation of the basic design for different institutions increased the complexity and engineering cost, it enabled them to follow different R\&D paths while still benefiting significantly from the common framework.

One area that was not considered sufficiently at the design phase was defining a suitable dimensioning scheme to mount the vacuum system in the packaging frames. The packaging frame was essentially a cuboid and provided a reasonable set of datums. However, the location of the parts of the vacuum system was not defined with respect to this frame, and it was not checked that the assembly tolerances were compatible. The lack of consistent dimensioning led to the vacuum chambers having to be tweaked into place when the whole system was almost complete, and success was not guaranteed until the final adjustments were made. In future such a system should recognise that, whereas there need not be any requirement on how the vacuum system sat in the frame, there is a need to specify positions to facilitate accurate assembly.

\subsubsection{Processing and assembly of the vacuum systems} \label{sec:Processing_Vacuum_System}

The AION sidearms have a modular design similar to accelerators and hence lend themselves to Daresbury Laboratory Engineering Technology Centre facility's construction processes. As an STFC National Laboratory facility, this Centre is set up to deliver `plug and play’ modules to major national and international accelerator projects. Although the instrument is predominately a vacuum system, the AION project called on the Centre's in-house mechanical workshop for modifications to components and the electrical technicians to provide power to the ion pumps during transportation and control of the strontium oven.  Rigging effort was also used to move systems and crates for delivery.  Having these teams under one roof kept the build progressing in a timely and efficient manner.

The bulk of the AION build was conducted through the Centre's vacuum processing laboratory, where high-precision cleaning and temperature processes are used to reach UHV specifications such as those required by the AION project and many accelerators.  The facility utilises a scientifically-designed cleaning process to prepare vacuum components to achieve specified vacuum levels.  The process uses a high-pressure detergent rinse and ultrasonic solvent cleaning to remove any hydrocarbons from the surface. Some components were then vacuum fired to drive hydrogen out of the stainless steel to reduce outgassing in order to improve the results of the final system bakeout.  The capacity of the facility made it possible to process the components of several systems in a minimal number of batches before assembly. As the laboratory routinely handles vacuum systems, it had a stock of suitable adaptors, spares and consumables that allowed the work to progress without unnecessary delays.

The subsystems were assembled and helium leak-checked to resolve any leaks before initial residual gas analysis (RGA) and temperature bakes of components.  This highlighted a significant weakness in the supplied commercial viewport windows and several initially failed the leak rate required for the UHV systems and others were not robust enough for a heat cycle.  One of the specified coatings did not adhere well to the viewports and several seemed to deteriorate with heat cycling.  The project decided to reduce their number and shuffle viewports from other systems, something which was made possible by the centralised assembly.

The heat cycle chosen for the bakeout was aimed to reduce any thermal differences across viewports and remain below the maximum allowable for the viewports, using a \SI{5}{\degree C} per hour ramp rate and maximum of \SI{140}{\degree C}.  This maximum was reduced to \SI{115}{\degree C} when the only available replacement windows were bonded rather than brazed. ~\footnote{A failure of a bonded Oxford Chamber 3 window on the final bake meant this had to be swapped to a blank flange without a bake, hence the tabulated pressure is above specification for this chamber, see Table~\ref{tab:vacuum_pressure}.}  Due to the project timeline the decision was taken to ship to Oxford without an additional recovery bake: this will be completed onsite once a replacement viewport is sourced.  Optical viewports proved to be a major challenge during the assembly and commissioning, and having an inventory across several systems meant  replacements could be made quickly. Future projects should carry a number of spares.

\begin{figure}[t]
    \centering
    \includegraphics[width = 0.95\textwidth]{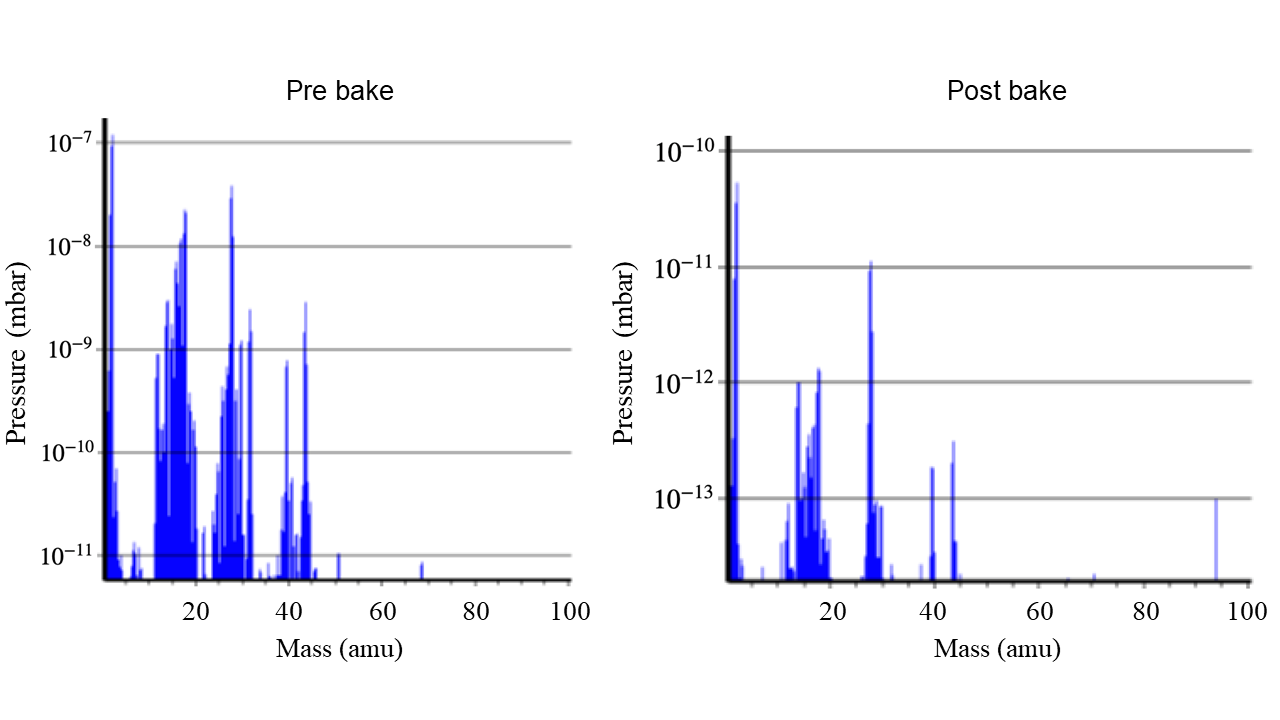}
    \caption{RGA scans \textit{(left)} pre- and \textit{(right)} post-bake, indicating the cleaning of the system: note the difference between the pressure scales.
        Data measured by the Daresbury Laboratory team.}
    \label{fig:RGA}
\end{figure}

Once the subassemblies had been proven to be clean and leak-tight to the appropriate level, the whole system was assembled into the frame and loaded into a large air oven for full bakeout procedure and NEG activation.  Knowing that the components are clean and leak-tight before assembly into the frame reduced the risk of a major strip down to resolve problems.  The system was connected to the oven turbomolecular pump, an RGA, leak detection and bakeout were performed.  Part-way through the bake the NEG pumps were activated to allow the gas released to be pumped away to reduce chamber contamination.  The ion pumps were flashed during system cool down.  Once complete, the system was leak checked, another RGA scan was taken and any issues resolved. Comparing the pressures found in RGA scans before and after baking, shown in the left and right panels of Fig.~\ref{fig:RGA},
very substantial reductions in the residual gas pressures over a wide range of atomic mass units (AMU) can be seen.  Furthermore, the centralised production approach led to an improved knowledge of expected contaminants before and after bake, and an understanding of which of these contaminants would cause vacuum issues.  The scans in Fig.~\ref{fig:RGA} show no presence of hydrocarbons prior to bake and the dominant pre-bake water contaminant (AMU 18) was greatly reduced by the bakeout.

The final step was to load the strontium oven in an inert atmosphere followed by loading into the vacuum system under nitrogen purge.  The vent and pump down at this point was kept to a minimum time to reduce ingress of water into the system and recover a good vacuum level.  The system was then taken through a final bake cycle where the strontium oven was outgassed and run up to temperature.  After final checks were complete the system was sealed and removed from the bake oven on the transport trolley ready for packing into the crate for transport to the institutes. The vacuum pressures achieved in each of the sidearm chambers are shown in Table~\ref{tab:vacuum_pressure}.

The fact that the facility had responsibility for several builds permitted a certain amount of parallelisation.  This became beneficial when a design problem was met that had to be resolved, as work could continue on subsequent builds.  As the basic design was common the resolution was applied uniformly across all systems.  The learning gained from the first system allowed the following ones to progress at a much faster rate.  Several challenges related to the assembly from the design could be improved in subsequent iterations of the system.

\begin{table}[]
    \centering
    \begin{tabular}{p{3cm}ccc}
        \toprule
                             & \multicolumn{3}{c}{\textbf{Pressure (mbar)}}                                            \\
        \textbf{Institution} & \textbf{Chamber 1 }                          & \textbf{Chamber 2 } & \textbf{Chamber 3} \\
        \midrule
        Birmingham           & \num{5.7e-10}                                & \num{1.8e-11}       & --                 \\
        Cambridge            & \num{4.7e-10}                                & \num{2.0e-11}       & \num{2.1e-11}      \\
        Imperial             & \num{3.4e-10}                                & \num{1.3e-11}       & --                 \\
        Oxford               & \num[]{2.3e-10}                              & \num[]{1.6e-10}     & \num[]{2.1e-10}    \\
        RAL                  & \num{2.1e-10}                                & \num{1.3e-11}       & --                 \\
        \bottomrule
    \end{tabular}
    \caption{Steady-state pressures of each of the chambers that form the sidearm after baking and several days of operation, listed by institution. Pressures are judged by the currents in the ion pumps linked to the chambers. In view of the conductance of the pump linkages, the pressures experienced by the atoms may be slightly higher. Note that the pressure in the University of Oxford Chambers 2 and 3  is higher than specification due to the replacement of a failed window and no recovery bake: see the text for more details.}
    \label{tab:vacuum_pressure}
\end{table}
\vspace{10mm}

\subsubsection{Logistics and installation planning}
\label{sec:logistics}
\begin{figure}[t]
    \centering
    \includegraphics[width = 0.8\textwidth]{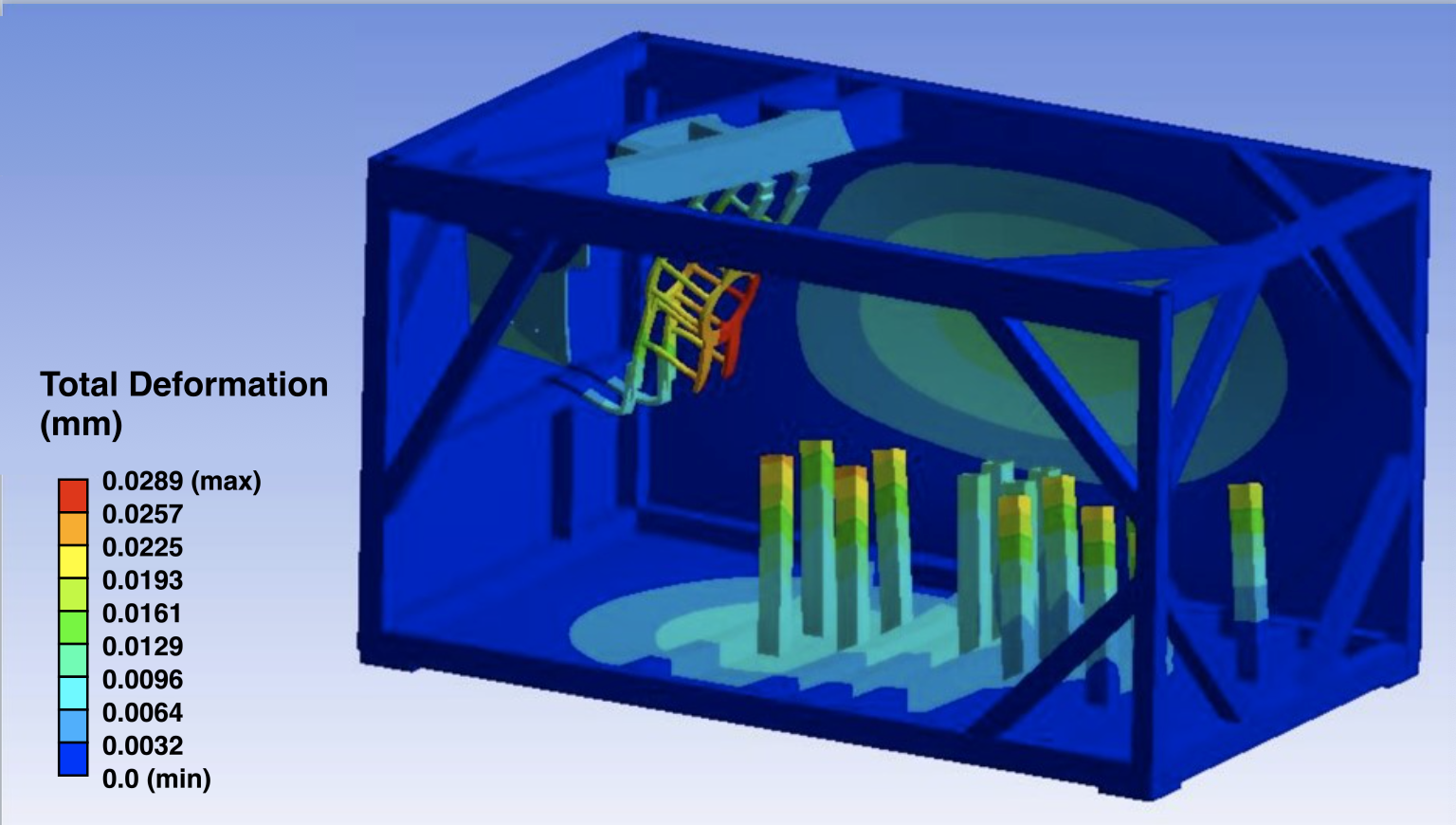}
    \caption{Results for total displacements found in a Finite Element Analysis (FEA) of the transport package structure, illustrating the combined use of space frame and shear plates to support the vacuum systems during transport.}
    \label{fig:FEA_analysis}
\end{figure}

\begin{figure}[t]
    \centering
    \includegraphics[width = 0.5\textwidth]{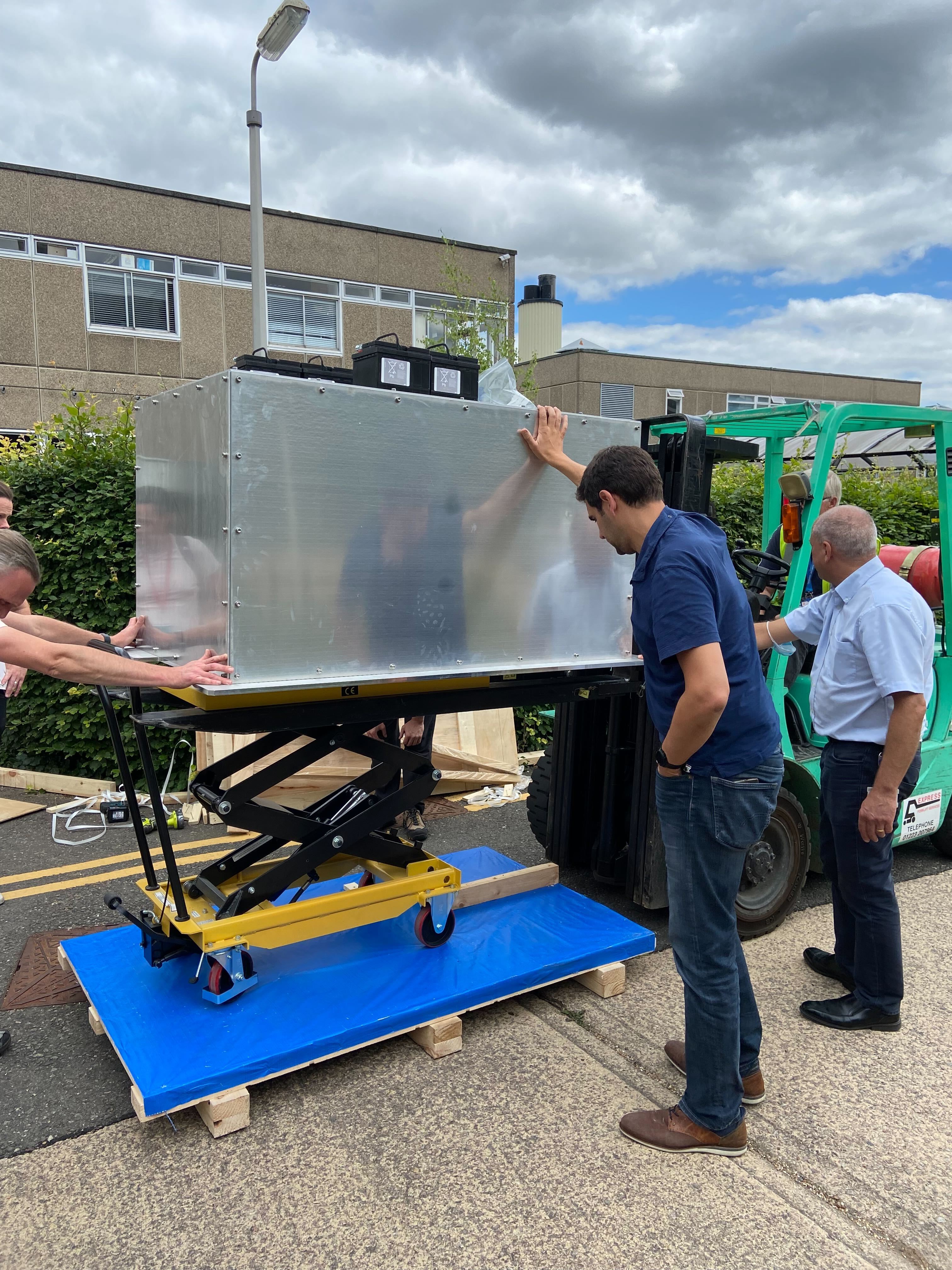}
    \caption{The University of Cambridge sidearm being unloaded with shearplates in place - showing the transport and assembly trolley, and the base of the wooden shipping case.}
    \label{fig:Transport_trolley}
\end{figure}

The design of the transport process included vibration simulations on the sidearm with representative transport vibration spectra to ensure that no part of the system would be displaced by a problematic amount. Finite Element Method (FEM) studies were used for this purpose, and showed the diagonal stiffness of the frame to be marginal in some cases.
Therefore, the assembled vacuum system in the sidearm framework was wrapped in shear plates (6~mm thick aluminium sheet metal to provide at minimal cost substantial diagonalisation of the frame's cuboid faces) to improve the framework's marginally adequate racking stiffness.
Fig.~\ref{fig:FEA_analysis} shows results from the displacement study that motivated this choice.
The enclosed sidearm was then packed in a custom wooden crate along with the trolley it was built upon and a power system to keep the ion pumps running throughout the transport (see Fig. \ref{fig:Transport_trolley}).

\begin{figure}[tb]
    \equalSubFigs{fig8/Sidearm_on_baseplate.jpeg}{fig8/sidearm_by_table.jpeg}[\textwidth]
    \caption{
        \textit{(Left)} The RAL sidearm on its transport baseplate following the removal of the surrounding shearplates. The batteries used to power the vacuum pumps during transport can be seen on the left of the image.
        \textit{(Right)} Movement of the RAL sidearm from the transport trolley to the optical table using translating rails (see text).
    }
    \label{fig:Sidearm_install}
\end{figure}

Getting the sidearm off the transport trolley could not be done by lifting as none of the destination laboratories had either built-in lifting equipment or the ability to bring in a temporary system. Instead a system of rails was designed that could be used to bridge between the transport trolley and the designated optical table, Fig.~\ref{fig:Sidearm_install}, and the sidearm had castors attached that were used to jack the sidearm up and roll it into position. In some instances multiple moves were required to position the sidearm correctly in the laboratory.

\begin{figure}[t]
    \centering
    \includegraphics[width = \textwidth]{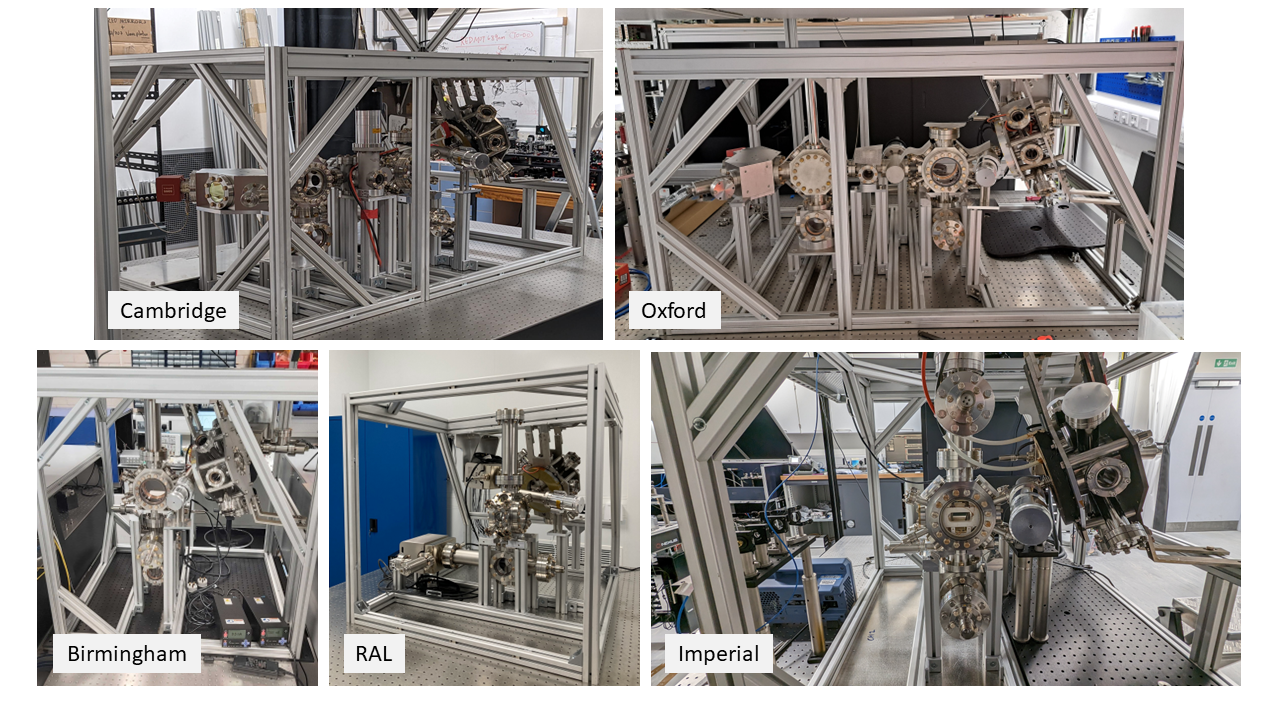}
    \caption{The five sidearm systems, installed at their corresponding institutions.}
    \label{fig:all_sidearms}
\end{figure}

Once the sidearm was placed, it was bolted down to the table and the rest of the transport framework could be dismantled from around the vacuum system. Figure~\ref{fig:all_sidearms} shows the five sidearms installed at their respective institutes. All sidearms were delivered and installed successfully without the move creating any vacuum failures. Although this transport and delivery process was well designed, a simple change during manufacture switched away from the original brackets to through-bolted connections, which made the removal of components very complex.

\clearpage

\subsection{Magnet System}
\label{sec:magnets}

\subsubsection{Magnetic field requirements}
\label{sec:field_requirements}

As described in Section~\ref{sec:laser_cooling_Sr}, the AION sidearm
requires control of magnetic fields and field gradients in two locations, Chambers 1 and 2, in order to facilitate the 2D MOT in Chamber 1 and the ``blue" and ``red" 3D MOTs in Chamber 2. In particular they require a 2D quadrupole field in Chamber 1 and a 3D quadrupole in Chamber 2. In order to estimate the required gradients, the {\tt AtomECS} Rust package~\cite{Chen2021atomecs} was used to simulate the laser cooling dynamics with the goal of optimising atom capture rates in the 3D ``blue" MOT in Chamber 2. The
{\tt MagPyLib} Python library~\cite{magpylib} was used to determine how the required field strength could be obtained in Chamber 1, and the Chamber 2 magnets were simulated in {\tt Mathematica} using the Radia package.~\footnote{
{\tt https://www.esrf.fr/Accelerators/Groups/InsertionDevices/Software/Radia}}

\subsubsection{Magnet design - Chamber 1}
\label{sec:design1}

\begin{figure}[tb]
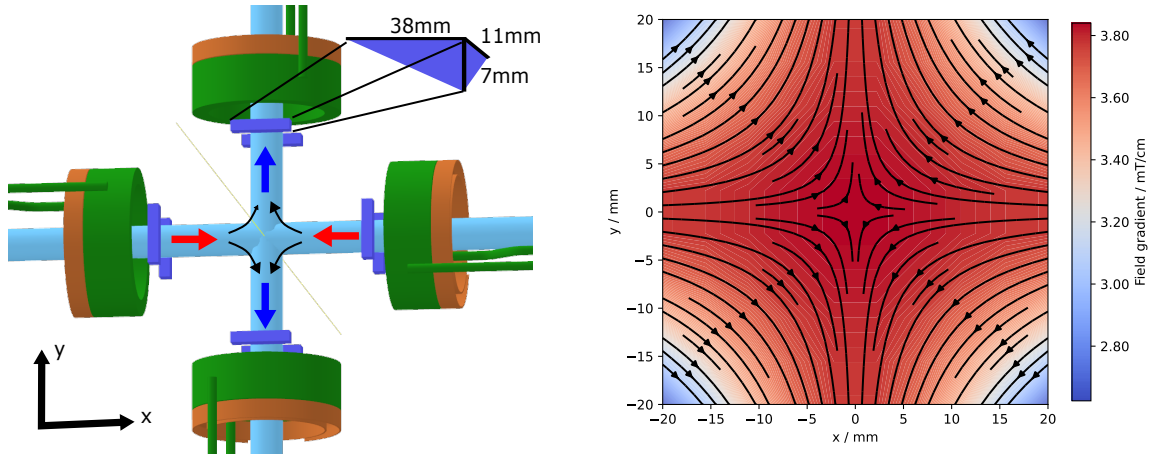

    \equalSubFigs{fig10/ch1_magnets.pdf}{fig10/2d_quad_field.pdf}[\textwidth]
    \caption{
        \textit{(Left)} Chamber 1 coil assemblies, showing permanent magnets in dark blue, trim coils in green, and heat exchangers in orange. Red (blue) arrows indicate the north (south) poles of the magnets. 
        Doppler-cooling laser beams (light blue) counter-propagate along the $x$ and $y$ axes. Atoms experience cooling and confinement in the $xy$ plane, and are pushed in the $z$ direction by an additional beam into the second chamber.
        \textit{(Right)} Simulated magnetic field lines and gradients generated in Chamber 1 at the calculated optimum parameters for atom loading in Chamber~2.
    }
    \label{fig:Ch1_2D}
\end{figure}

Chamber 1 includes three separate magnet systems, see Fig.~\ref{fig:Ch1_2D}. 
Strong radial field gradients are provided by four assemblies of pairs of N52 NdFeB permanent magnets.
Using a simplified model of the system, the simulated magnetic fields of these eight permanent magnets were shown to create a quadrupole magnetic field in its centre, and the {\tt AtomECS} Rust package~\cite{Chen2021atomecs} was used to optimise the atomic flux by varying the dimensions and distance of the Chamber 1 permanent magnets.

The optimum was obtained with the N52 magnets orientated perpendicular to the 2D MOT laser beams, located with their closest face coincident with a plane
\SI{65.1}{\milli\meter} from the centre of the chamber, their base offset by
\SI{20.5}{\milli\meter} from the laser beams, with dimensions
\qtyproduct[product-units = single]{11 x 38 x 7}{\milli\meter}, and with their
largest faces orientated towards the atoms as shown in Fig.~\ref{fig:Ch1_2D}.
This produces a field gradient of
\SI{3.7}{\milli\tesla\per\centi\meter} on each axis in the $xy$ plane at the centre of the chamber,
and the calculated field at all locations was used in the {\tt AtomECS}
atom-capture simulation.

Rotating the magnets such that they faced towards the atoms would have resulted
in a greater field gradient per unit magnet volume, with the optimum angle
calculated as around \SI{20}{\degree}. However, use of this angle would only
have reduced the magnet volume by \SI{10}{\percent} with a similar reduction in
the stray field gradients introduced in Chamber 2. Given the increased design
complexity of machining angled magnet holders, it was decided to adopt perpendicular
permanent magnets.

The simulated rate of atom loading was found to be approximately equal for
magnet-to-atom distances varying by $\pm~\SI{5}{\milli\meter}$.
Our design also includes paired solenoid coils that allow for trimming of the field gradient
for later empirical optimisation, and also for the application of bias magnetic
fields in order to control the location of the 2D MOT - an important parameter
since the atoms are then launched through a \SI{3}{\milli\meter} diameter differential
aperture between the two chambers.

The trimming coils consist of four water-cooled aluminium tape
coils made of \SI{30}{\milli\meter} wide and \SI{0.2}{\milli\meter} thick anodised
aluminium tape, supplied commercially as a custom design by Anoxal~\cite{anoxal}.
The use of aluminium tape resulted in compact, robust coils that self-insulate
through anodisation of the aluminium layers, despite supporting current
densities of up to \SI{10}{\ampere\per\milli\meter\squared}. Aluminium has a lower
electrical conductivity than copper, but the constrained space meant that
overall volume drove our design requirements. The high thermal conductivity of
aluminium combined with no need for electrical insulation between coil layers
permitted cooling from a single contact point, away from the atoms, via a copper
heat-exchanger (visible in Fig.~\ref{fig:Ch1_2D}) coupled via a
\SI{3}{\watt\per\meter\per\kelvin} silicon pad.

After the iterative field and atom dynamics simulations had been performed, a
final simulation was made using a commercial FEM
software package, to confirm agreement with the simplified model of the permanent magnet
fields and to characterise the additional effect of the trimming coils.
Good agreement between the models was found. The trim coils should supply
up to an additional \SI{2.4}{\milli\tesla\per\centi\meter} if operated at their
maximum capacity, allowing for a large range of adjustment from the baseline
provided by the permanent magnets.

In addition, axial
compensation fields are applied through similar, tape-wound coils
(\SI{24}{\milli\meter}) from the same supplier. Since they provide only
bias field compensation requiring currents $<$~\SI{5}{\ampere}, passive cooling
sufficed.

\begin{figure}[tb]
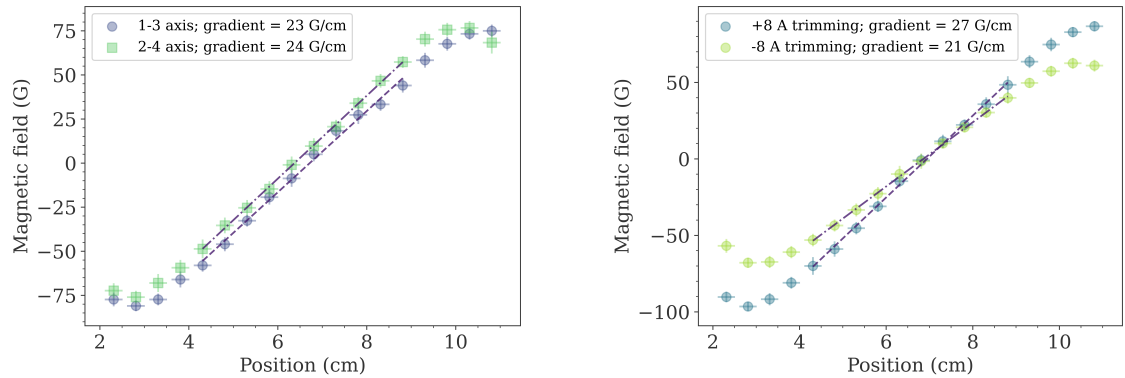

    \equalSubFigs{fig11/PermanentMagnets.png}{fig11/TrimmingCoils.png}[\textwidth]
    \caption{
        \textit{(Left)} Magnetic field in Chamber 1 generated along two perpendicular axes by two pairs of permanent magnets, arranged in the design configuration. The magnetic field gradient in the centre of the chamber is slightly less than the 27 G/cm predicted by simulations, and has been found to vary by up to 2 G/cm depending on the individual installation.
        \textit{(Right)} Magnetic field in Chamber 1 generated along the `1-3' axis by the design configuration of two pairs of permanent magnets and trimming coils, with a trimming current of 8 A applied to all coils. Coils can be tuned independently to modify the field gradient and zero-field location. Data measured by the University of Cambridge team.
    }
    \label{fig:Ch1_Cam}
\end{figure}

By minimising the volume requirements of the Chamber 1 coils, it was possible to
place permanent magnets housed in aluminium casings close to the atoms, reducing
their required volume and therefore minimising the effect of stray fields in
Chambers 2 and 3.
The magnet arrangement around Chamber 1 and the
resultant magnetic field profile are shown in Fig.~\ref{fig:Ch1_2D}. The field measured along one axis from two magnet pairs is shown in Fig.~\ref{fig:Ch1_Cam}: the dashed and dash-dotted lines are linear fits to the data over a range of positions including that required for operation.
The difference between the fits illustrates the effect of applying a current to a trimming coil.

\subsubsection{Magnet design - Chamber 2}
\label{sec:design2}

\begin{figure}[tb]
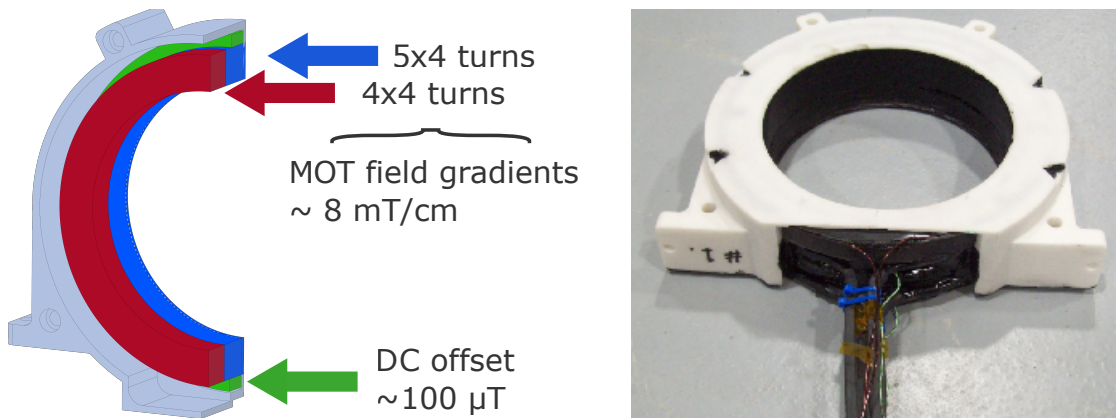

    \equalSubFigs{fig12/ch2_coils.pdf}{fig12/ch2_coils_real.png}[\textwidth]
    \caption{
        \textit{(Left)} Chamber 2 copper coil body showing the three individual coils involved. The red and blue coils are water-cooled in parallel but electrically connected in series and are capable of supporting \SI{200}{\ampere} of current flow indefinitely.
        \textit{(Right)} A completed Chamber 2 coil assembly, showing the 3D-printed resin former and Stycast potting. The semi-rigid current leads were intentionally left long so that water connections could be made far from sensitive optics to lower the risk of water damage.
    }
    \label{fig:ch2_coils}
\end{figure}

In Chamber 2, all fields are produced through electromagnets, making it possible to control the gradient over a large range and to ramp the total field to zero. The two Chamber 2 magnet assemblies
each contain two coils of square-profile, \qtyproduct{5x5}{\milli\meter} hollow copper conductor with \SI{3}{\milli\meter} internal apertures, cooled by internal water flow. These were wound in two parts as $5\times4$- and $4\times4$-turn coils embedded into the same housing - see Fig.~\ref{fig:ch2_coils}. The conductor was first insulated with heat-shrink PVC tubing before being wet-wound with Stycast resin.

The coils are electrically connected in series and the assemblies are run in
anti-Helmholtz configuration. Cooling water is run in parallel through
the four coils to reduce pressure requirements. An auxiliary,
passively-cooled winding was placed around the outer diameter of the coil
closest to the atoms to provide DC offset field trimming, and thermocouples were
embedded for interlocking and monitoring.

\begin{figure}
    \begin{center}
        \includegraphics[width = 0.7\textwidth]{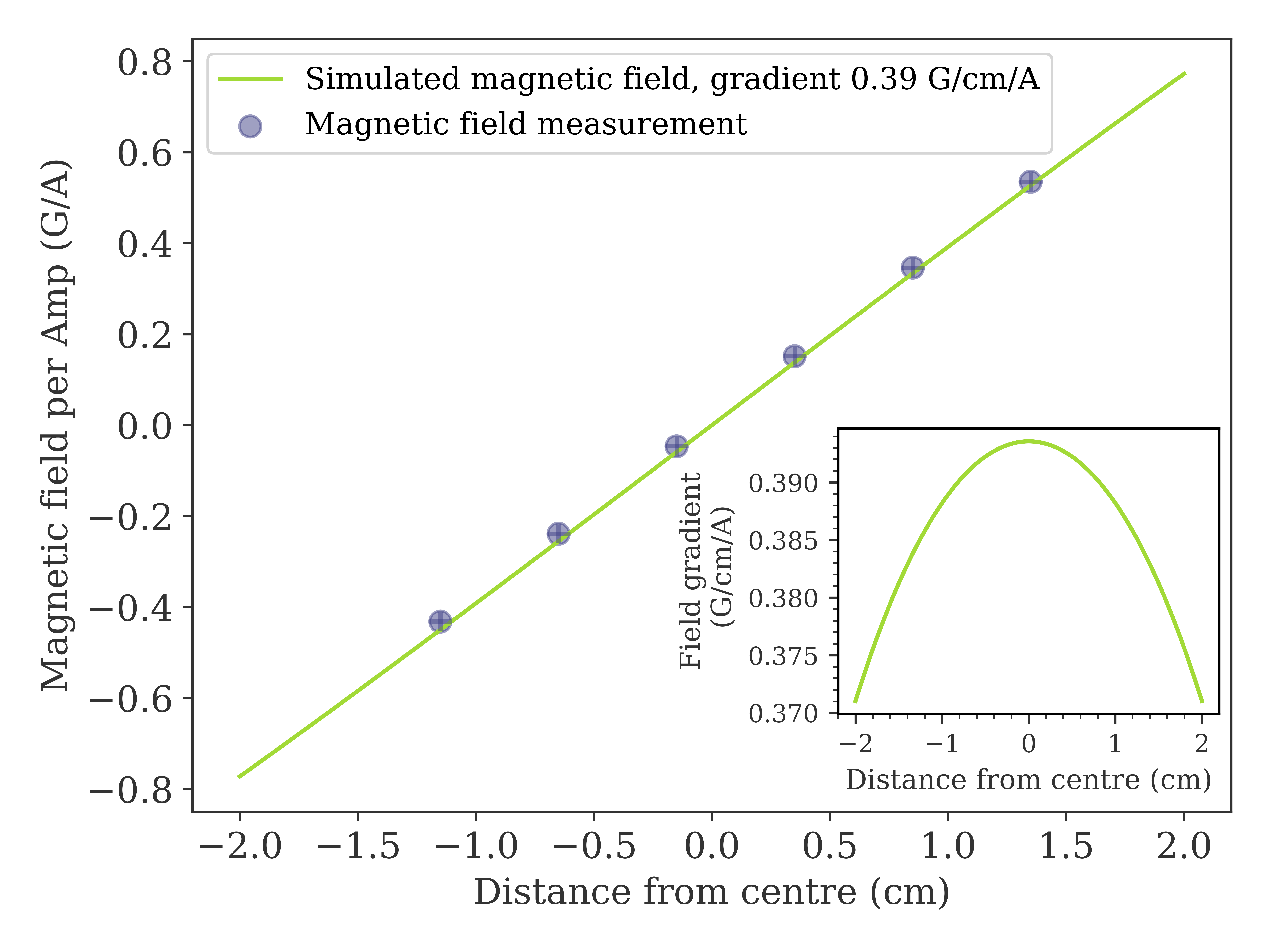}
        \caption{The axial magnetic field in Chamber 2 generated by two pairs of electromagnets, arranged in the design configuration. (Measured at a current of 10 A.) The measured data points indicate a gradient in agreement with that of a simulated field created by electromagnets in the design configuration with a layer of insulation 0.6~mm thick surrounding each wire. The inset shows the simulated magnetic field gradient in the centre of the coil array. Data measured by the University of Cambridge team.}
        \label{fig:Ch2_Cam}
    \end{center}
\end{figure}

Fig.~\ref{fig:Ch2_Cam} shows the axial magnetic field in Chamber 2 generated by the magnet assemblies in the design configuration. The field gradient of 0.39~G/cm/A in the trap centre is in good general agreement with field simulations assuming an insulating layer 0.6~mm thick surrounding each wire. Deviations from this simulation can be explained by inhomogeneities in the coil winding and casting, and tolerances in the installation configuration. With water cooling, currents of up to 200~A have been continuously applied without the coils exceeding a temperature of \SI{30}{\degreeCelsius}.

The self-inductance of the pair of coils installed in series and in anti-Helmholtz configuration around the chamber is $<$\SI{400}{\micro\henry}, allowing the current to be ramped within 1~ms from around 130~A to 10~A, as required for the transition from blue MOT to red MOT, at a reverse voltage of 40~V.

\section{Laser-Stabilisation System}
\label{sec:lasers}

\subsection{Motivation}
\label{sec:motivation}
The laser requirements were briefly outlined in Section~\ref{sec:laser_cooling_Sr}, and a level scheme including all the required laser wavelengths is shown in Fig.~\ref{fig:laser_cooling_diagrams}. The lasers at 461~nm (blue MOT), 679~nm and 707~nm (repumping) only require stabilisation at the \SI{1}{\mega\hertz} level. This stabilisation was implemented using a software-based PID lock to a commercial wavelength meter, the High Finesse WS8.

Two of the lasers---those at 689~nm (red MOT) and 698~nm (atom interferometry)---have stringent frequency stabilisation requirements that are the focus of this Section. In order to create consistent atom samples in the narrowband stirred red MOT \cite{mukaiyama_recoil-limited_2003}, and in order to match consistently the MOT position to the location of the dipole trap \cite{stellmer_production_2013}, the 689~nm laser must be stabilised at roughly the \SI{2}{\kilo\hertz} level or better. The 698~nm laser must be stabilised at roughly the \SI{10}{\hertz} level, in order to support high-fidelity $\pi$-pulses on the clock transition as needed for large momentum transfers \cite{chiow_102ensuremathhbark_2011,Rudolph:2019vcv}.

\subsection{Design}
\label{sec:design}
\begin{figure}[t]
    \centering
    \includegraphics[width = 0.9\textwidth]{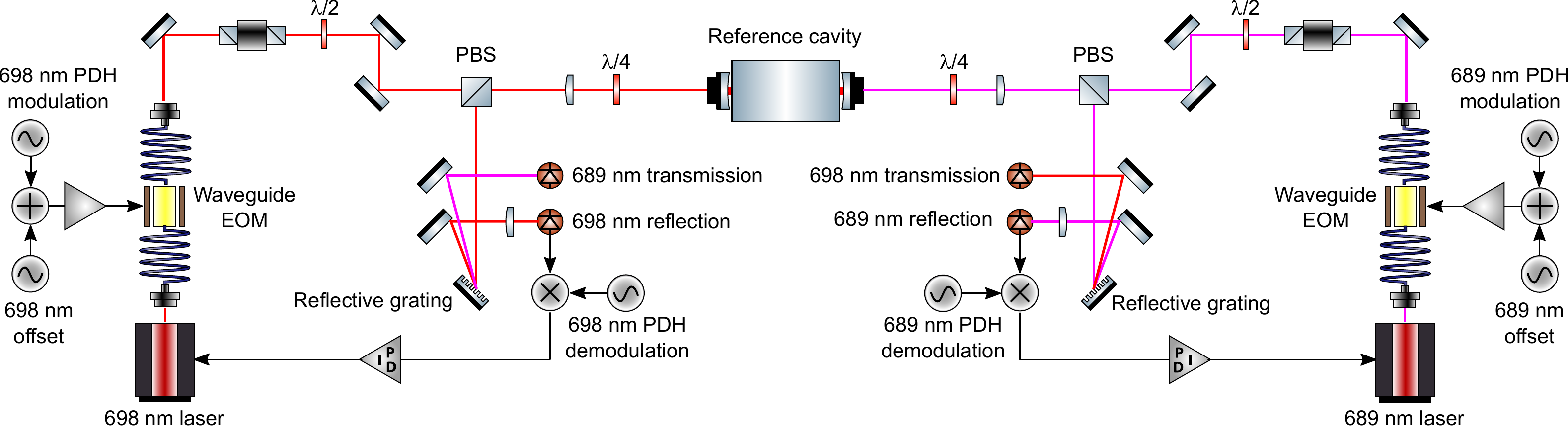}
    \caption{Optical layout for locking the 689~nm and 698~nm lasers to an optical reference cavity, using a dual-sideband PDH lock \cite{thorpe_laser_2008}. }
    \label{fig:Cavity_optics}
\end{figure}

Informed by these requirements, a system to frequency-stabilise both the \SI{689}{\nano\meter} and \SI{698}{\nano\meter} light was designed. The fundamental aspects of the system are shown in Fig.~\ref{fig:Cavity_optics}. The \SI{689}{\nano\meter} and \SI{698}{\nano\meter} beams propagate in opposing directions through a notched cavity \cite{webster_vibration_2007}, purchased from Stable Laser Systems~\cite{SLS_website}. The transmitted and reflected wavelengths from each laser are then separated onto different photodetectors. 

\subsubsection{Description of cavity system, including optics}
\label{sec:cavity}

The optical layout used for laser stabilisation is depicted in Fig.~\ref{fig:Cavity_optics}. A fraction of the light is split off from each of the main lasers at 689~nm and 698~nm and passed through a fibre-pigtailed waveguide EOM (Jenoptik PM705). This light then exits the fibre and is collimated by an aspheric lens. An optical isolator is used to avoid the reflection of light back to the main laser, and to suppress parasitic etalons between the cavity and the optical fibre tips. Following this, the light passes through a polarising beamsplitter cube (PBS) and a quarter-wave plate. Any non-resonant light that is reflected from the cavity then passes through the same quarter-wave plate again, and the light exits the PBS using the other, orthogonal, port. Diffraction gratings are used to separate spatially the reflected beams at one wavelength from the transmitted beam at the other wavelength ---a dichroic beamsplitter would be mechanically more convenient, but an appropriate optic was difficult to source because the laser wavelengths are close to each other. The transmitted beam provides a discriminator signal that helps relock the laser frequency following any disturbances.

\subsubsection{Pound-Drever-Hall method and reference}
\label{sec:PDH}
To generate an error signal with which the laser frequency is stabilised, the Pound-Drever-Hall (PDH) method is used~\cite{drever_laser_1983}. This technique is widely employed in precision measurement, spectroscopy, and optical communications. It involves feeding back a portion of the laser light that has been reflected from the reference cavity to the laser's frequency control system. In the AION system, a fibre-based electro-optic modulator (EOM) is used to modulate the laser light at around \SI{10}{\mega\hertz}. Upon interaction with the cavity this light is then converted to an electrical signal on a fast photodiode (Koheron PDX10S-5-DC-SI), which is then mixed with a local oscillator, generating an error signal proportional to the frequency difference between the laser and the cavity. The error signal is fed back to the laser frequency control system, which adjusts the laser frequency to reduce the error signal.

In order to lock the laser at a frequency corresponding to the atomic transition, which can be as much as half of the free spectral range (FSR), $\omega_\mathrm{FSR}/2 = 2 \pi \times \SI{750}{\mega\hertz}$, away from the nearest (randomly-positioned) cavity mode, a dual sideband lock is applied \cite{thorpe_laser_2008}. In this method, sidebands are created by driving the EOM with an offset frequency $\omega_\mathrm{offset}$ that is much larger than the PDH modulation frequency  $\omega_\mathrm{PDH}$. The laser is tuned and locked such that the frequency of the first-order sideband at $\omega_\mathrm{laser} \pm \omega_\mathrm{offset}$ is resonant with the cavity.
With the offset frequency's amplitude chosen appropriately to suppress the carrier component of the laser field, this method results in two possible lock-points within each $\omega_\text{FSR}$ frequency range of the cavity, separated by $\omega_\mathrm{offset}$, only one of which is desirable. Since $\omega_\mathrm{offset}$ is typically several hundred megaHertz, a commercial wavemeter is able to disambiguate the modes when the laser requires initial locking.

\subsubsection{Parallel construction and characterisation}
\label{sec:parallel}

Due to their centralised construction, the cavities could be compared with each other, to diagnose characteristics such as cavity mode linewidth, locked-laser linewidth, and short-term frequency instability. 
These comparisons could be done by locking separate lasers to their own cavity and performing a heterodyne measurement between the lasers. However, an alternative method is to use just one laser, combined with the dual sideband generation technique. The first cavity is used to lock the laser frequency directly to the cavity mode without offset, as described in Section~\ref{sec:PDH}. Then, a  fraction of this light is brought into resonance with a second cavity. This is done by adding a second frequency to the EOM,
which supplements the modulation used to derive the PDH signal. As the free spectral range of the cavity is \SI{1.5}{GHz}, this again means that the additional frequency may need to be anything up to \SI{750}{\mega\hertz} in order to bridge the gap. To account for the cavity mode drift, this additional frequency needs to be tracked and counted. In the AION implementation, the relatively slow drift of the cavities means that a low-bandwidth PID lock---which feeds back onto the EOM frequency---is sufficient. 

\begin{figure}[ht]
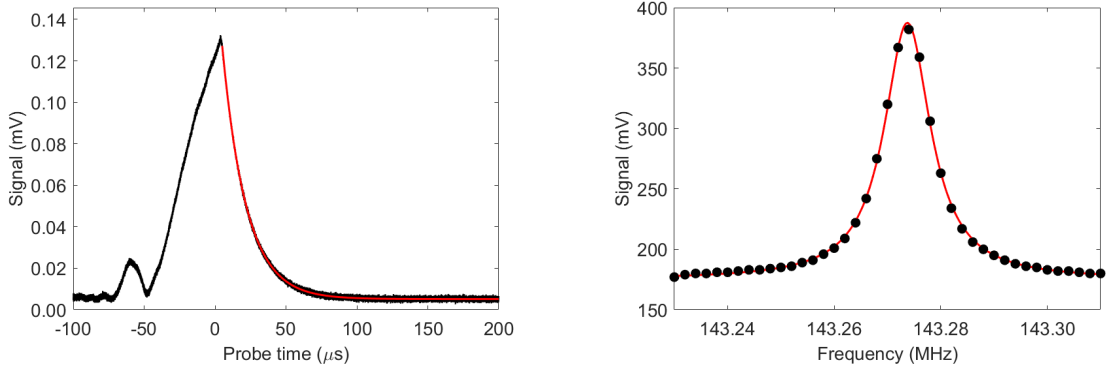

    \equalSubFigs[fig:Ringdown]{fig14/Ringdown2.png}[fig:Linewidth]{fig14/Linewidth_setAR.png}[\textwidth]

    \caption{
        \textit{(Left)} Cavity ringdown measurement at \SI{689}{\nano\meter}. The data are fitted with an exponential decay, giving a lifetime of \SI{16.9}{\micro\second}, implying a finesse of \num{160000}.
        \textit{(Right)} Cavity linewidth. The cavity linewidth is determined using a photodiode to measure light transmitted through the cavity as the laser frequency is scanned across the resonance. Fitting the transmission data gives a FWHM of \SI{10.6}{\kilo\hertz}. Data measured by the Rutherford-Appleton Laboratory team.
    }
    \label{fig:Ringdown}
\end{figure}

This comparison setup underpins the data presented in Fig. \ref{fig:Ringdown}, which show that the cavities have sufficient finesse to lock the lasers at the Hertz level. Further measurements with an RF spectrum analyser and frequency counter verified that the laser sidebands could be locked to separate cavities to a relative linewidth of \SI{10}{\hertz}, with a frequency instability of \SI{3}{\hertz} at 1 second averaging time.

\subsubsection{Conclusion and benefits of parallel construction of the laser stabilisation systems}
\label{sec:conclusion}

\begin{figure}[ht]
    \centering
    \includegraphics[width = 0.95\textwidth]{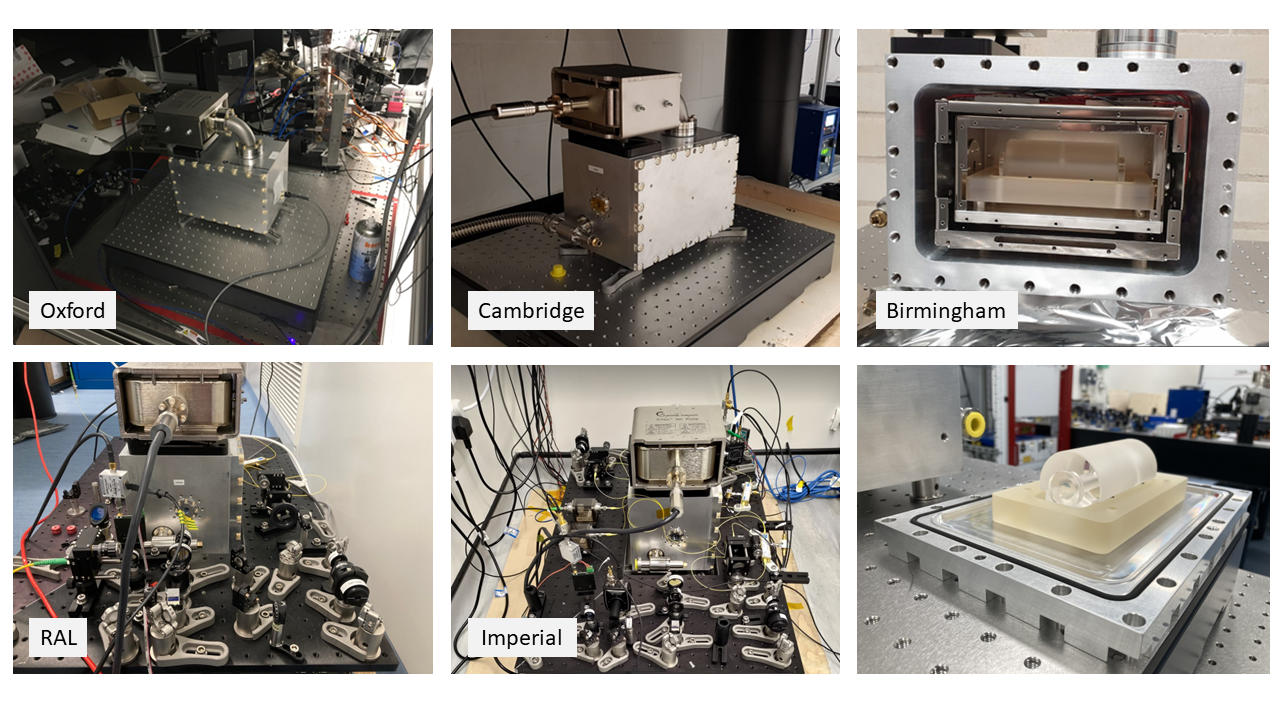}
    \caption{Cavity assemblies. The five cavity systems at various stages of construction. Also shown \textit{(lower right)} is a cavity prior to installation inside the vacuum housing. All optics were assembled centrally prior to delivery. Cavities were transported outside of their vacuum housing and reassembled locally at each institute.}
    \label{fig:Cavities}
\end{figure}

The laser stabilisation systems for all five strontium laboratories were built in parallel in one location, at RAL, before being distributed to the partner institutions, as shown in Fig.~\ref{fig:Cavities}. The approach of parallel construction had several advantages for the project.

The primary motivation, and subsequent benefit, of this approach was efficiency. By centralising the design and construction, an optimised, more streamlined process was developed, harnessing the expertise of team members with prior experience working with ultra-stable laser systems. By doing so, the systems could be constructed more efficiently and in a shorter time frame than had they been assembled locally at each institute.

With the laser stabilisation systems co-located, it was possible to characterise them against each other; a key advantage not afforded by local assembly. It is anticipated that in the future more efficient debugging, maintenance and operation of the laser stabilisation systems will be possible in the five collaborating laboratories, with rapid and transparent sharing of experience between them.

\section{Cold Atom Commissioning}
\label{sec:Beam_commissioning}

\subsection{Strontium oven}
\label{sec:oven}

\begin{figure}[ht]
\centering
\includegraphics[width = \textwidth]{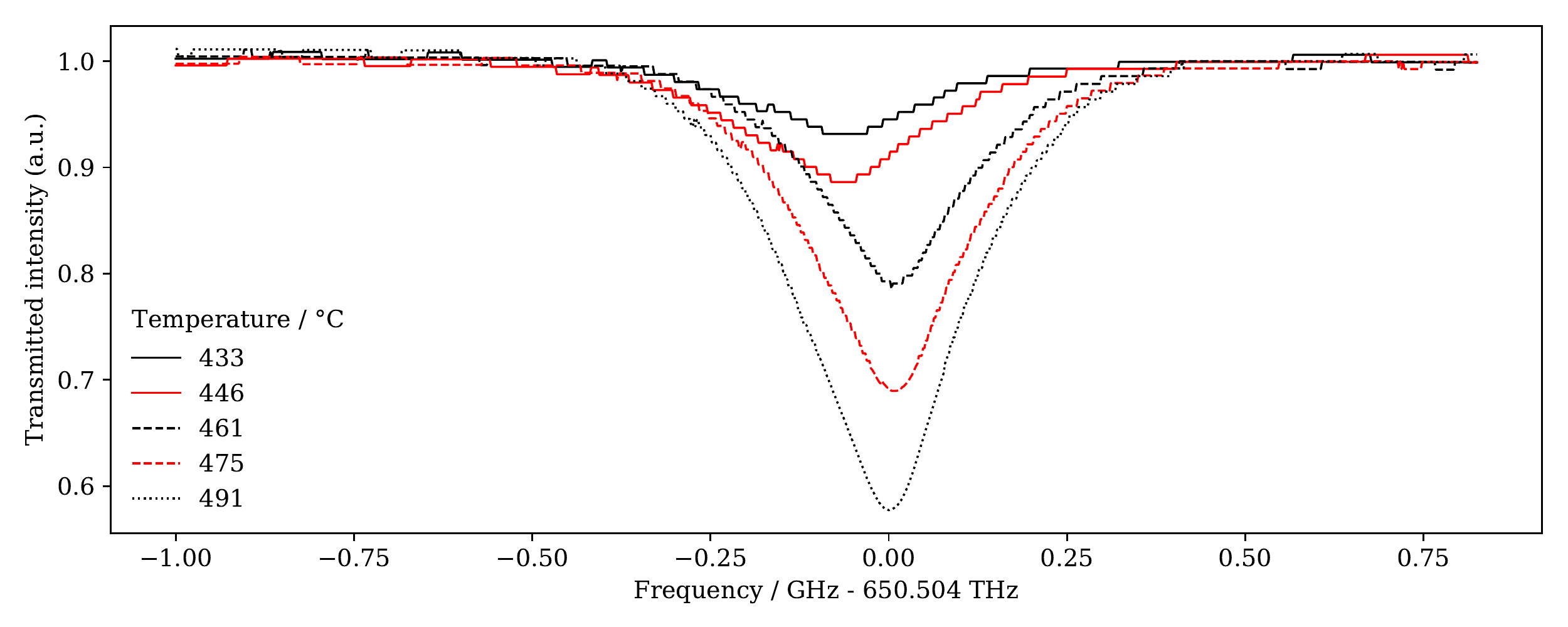}
    \caption{
        Absorption spectra of the atomic beam at various oven temperatures. The laser beam was normal to the atomic beam and the transverse velocity distribution is non-Gaussian for atoms effusing through the cylindrical channels in the nozzle (see Fig.~\ref{fig:oven}) of the oven~\cite{Greenland_1985}. Data measured by the Cambridge team.
    }
    \label{fig:OvenTemperatures}
\end{figure}

The sidearm design has been validated by demonstrating first-stage cooling of strontium in a 2D MOT in Chamber 1 and a first 3D ``blue" MOT in Chamber 2. As a first step, the atomic beam coming out of the oven was observed through fluorescence and absorption spectroscopy on the broad ``blue" transition, as seen in Fig.~\ref{fig:OvenTemperatures} for various oven temperatures.

For the first MOTs, the strontium oven was heated to \SI{420}{\degree C}, which raised the pressure in Chamber 1 from \SI{5.7e-10}{\milli\bar} at \SI{22}{\degree C} to \SI{6.5e-9}{\milli\bar}, as deduced from the ion-pump current. This oven temperature is low compared to the sublimation temperature of strontium and will be increased as the atomic flux of the sidearm is optimised. The oven design has been tested at temperatures of up to \SI{650}{\degree C}.

\subsection{2D ``blue" MOT}
\label{sec:2DMOT}

\begin{figure}[ht]
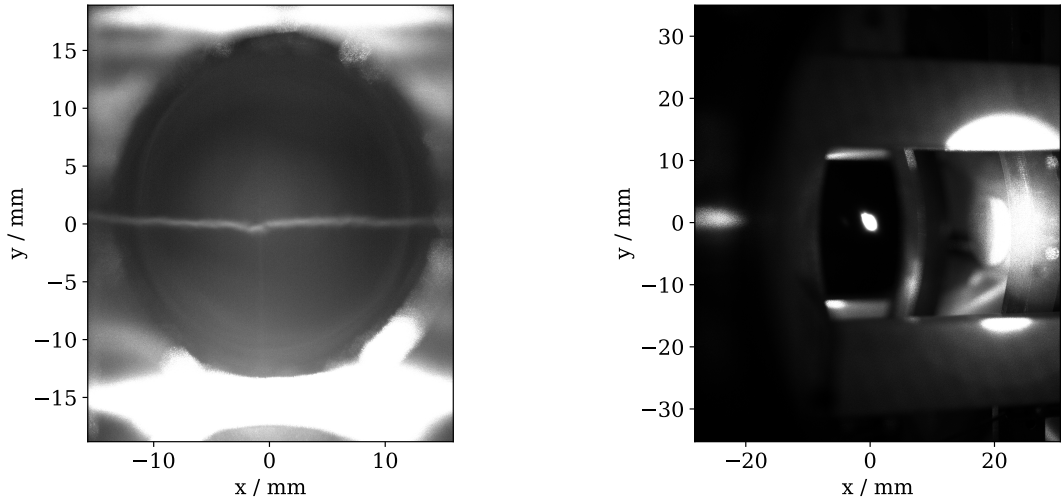

    \centering

    \equalSubFigs{fig16/2dmot.pdf}{fig16/3dmot.pdf}[\textwidth]

    \caption{%
        \textit{(Left)} Fluorescence from strontium atoms in the 2D ``blue" MOT in the Imperial sidearm's Chamber 1, viewed orthogonally to the push beam axis. Atoms are laser-cooled and compressed radially with minimal confinement along the longitudinal, push-beam axis, resulting in an elongated cloud visible as the thin, grey line stretching horizontally across the frame.
        \textit{(Right)} Image of a 3D ``blue" MOT in Chamber 2, captured from the atoms pushed by the 2D MOT in Chamber 1. The strontium cloud is visible as a small spot in the centre of the frame, and the surrounding structure is an optical cavity spacer used for creating squeezed quantum states that is unique to the Imperial sidearm variant.
        All spatial dimensions are given in the focal plane of the atom clouds. Images taken by the Imperial team.  
    }

    \label{fig:MOTs}
\end{figure}

In order to trap atoms in the 2D ``blue" MOT, light detuned by $-2\Gamma$ from the \textsuperscript{1}S\textsubscript{0}  to \textsuperscript{1}P\textsubscript{1} transition in \textsuperscript{88}Sr is used. Initial testing employed \textsuperscript{88}Sr, due to its natural abundance and bosonic nature, easing some of the frequency requirements of the laser system. Two circularly-polarised \SI{45}{\milli\meter} beams of intensity \SI{23\pm0.5}{\milli\watt\per\centi\meter\squared} and \SI{26\pm0.5}{\milli\watt\per\centi\meter\squared} were delivered into the 2D MOT chamber and
retro-reflected to create the four trapping beams. A 2D MOT of order \SI{20}{\milli\meter} in the untrapped dimension was observed in Chamber 1, see left panel of Fig.~\ref{fig:MOTs}. An atomic beam was created with the addition of a near-resonant weak push beam along the un-trapped axis to guide atoms from Chamber 1 to Chamber 2, via the \SI{3}{\milli\meter} differential pumping aperture. In this test, the magnetic fields for the 2D ``blue" MOT were supplied from the permanent magnets only, but they can be tuned in the future using the coils to optimise the atom flux through the system. 

\subsection{3D ``blue" MOT}
\label{sec:3DMOT}

An initial 3D MOT was loaded in Chamber 2  from the pre-cooled atomic beam from the 2D MOT, see right panel of Fig.~\ref{fig:MOTs}. 
For the 3D ``blue" MOT, three circularly-polarised orthogonal beams of \SI{15}{\milli\meter} at \SI{32}{\milli\watt\per\centi\meter\squared}, \SI{36}{\milli\watt\per\centi\meter\squared}, and \SI{18}{\milli\watt\per\centi\meter\squared}, were retro-reflected to create the six trapping beams, and detuned by $-2\Gamma$ from the \textsuperscript{1}S\textsubscript{0} to \textsuperscript{1}P\textsubscript{1} transition in \textsuperscript{88}Sr. 679~nm and 707~nm lasers (Fig.~\ref{fig:laser_cooling_diagrams}), resonant with their respective transitions, were used to pump back into the cooling cycle atoms that fall into the \textsuperscript{3}P\textsubscript{0} and \textsuperscript{3}P\textsubscript{2} states.
A current of \SI{125}{\ampere} was used in the quadrupole coils to provide a field gradient of \SI{4.9}{\milli\tesla\per\centi\meter}. 

\subsection{Atom flux measurement}
\label{sec:flux}

\begin{figure}[t]
    \centering
    \includegraphics[width = 0.9\textwidth]{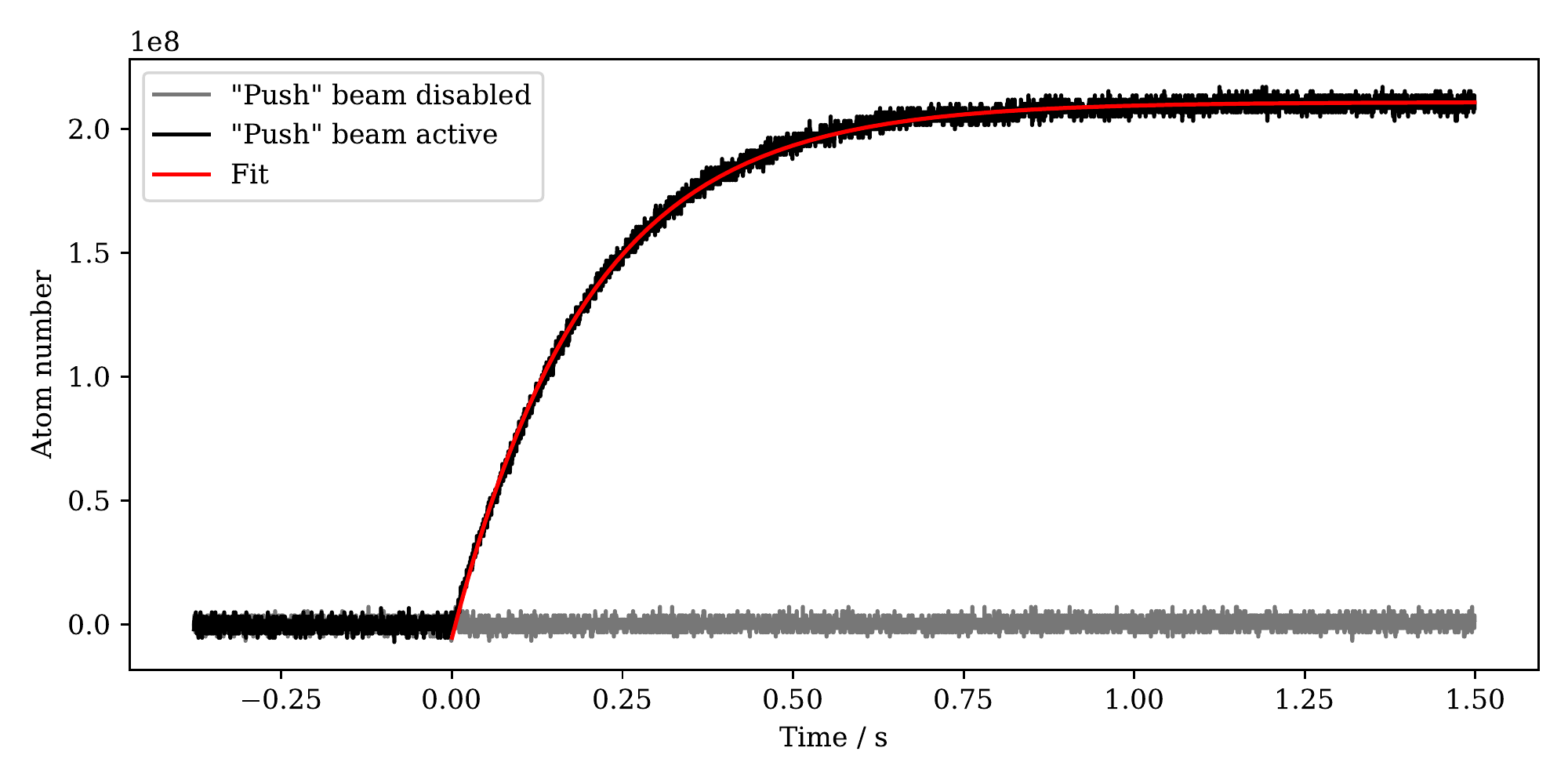}
    \caption{The 3D MOT loading curve from the 2D MOT atomic beam. MOTs of order 10\textsuperscript{8} atoms are loaded with time constant of \SI{198.8\pm 0.2}{\milli\second}, as measured by an exponential fit as in Eq.~(\ref{Loading}) that is
        represented by the red line. With the push beam off there is no loading seen from the 2D MOT alone, showing no diffusion through the \SI{3}{\milli\meter} differential pumping aperture.
        Data measured in the University of Birmingham sidearm.}
    \label{fig:LoadingCurve}
\end{figure}

The performance of the 2D ``blue" MOT can be assessed from the loading rate, $\tau$, and the equilibrium atom number, $N_{\rm eq}$, of the 3D ``blue" MOT in Chamber 2. The number of atoms, $N$, at a time, $t$, of the MOT is given by \cite{MOTloading}
\begin{equation}
    N(t) = N_{eq}(1-\exp(-t/\tau)).
    \label{Loading}
\end{equation}
Our preliminary results can be seen in Fig.~\ref{fig:LoadingCurve}, where the loading rate of the 3D MOT is shown for two atomic beam cases: with the 2D MOT and push beam on, and with just the 2D MOT light on in Chamber 1. In each case, the oven temperature was kept constant, all intra-vacuum valves remained open, and the 3D MOT light kept on. The 3D MOT coils were turned on at $t = 0$\,s to load the 3D MOT in each case. 10\textsuperscript{8} atoms were loaded into the 3D MOT trap, with a loading time constant of \SI{198.8\pm0.2}{\milli\second} estimated from the fit. The loading rate and atom number can be increased further by increasing the oven temperature to optimise the atom flux. In the cases where the push beam was shuttered, the loading rate of the 3D ``blue" MOT dropped to zero. This shows that leakage from the 2D MOT, and diffusion from Chamber 1 to Chamber 2 through the differential pumping aperture, is negligible. Loading from background vapour in Chamber 2 is also not seen, suggesting that the partial pressure of strontium in Chamber 2 is low.

\begin{figure}[t]
    \centering
    \includegraphics[width = \textwidth]{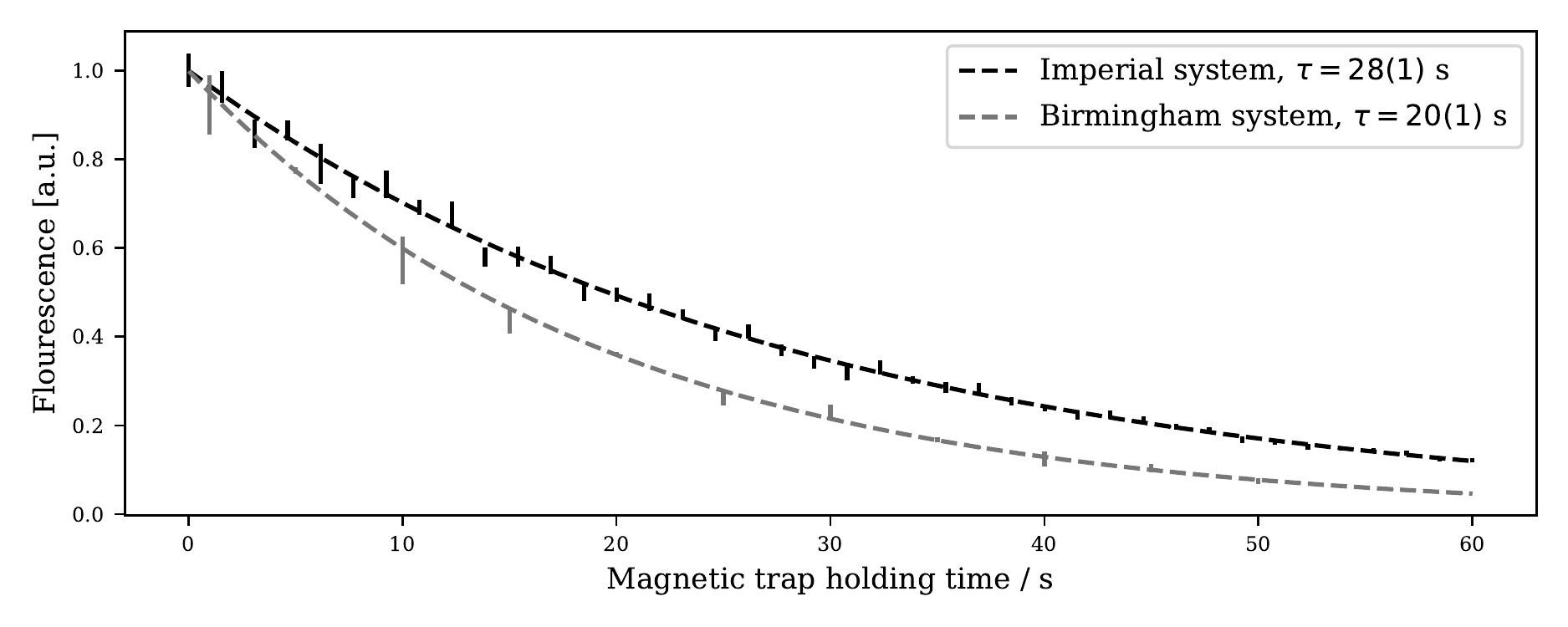}
    \caption{
        The lifetimes of the magnetic traps for atoms in the \ch{^3P_2} \ch{m_F}=2
        state in the Imperial and Birmingham sidearms.
        The figure shows the level of background-corrected fluorescence measured from atoms recaptured in the \ch{^1S0} to \ch{^1P1} MOT after holding atoms for a variable length of time in the magnetic trap. 
        Fitted time constants of \SI{28(1)}{\second} and \SI{20(1)}{\second} were measured from ten and four repeats of each holding time for the Imperial College and University of Birmingham sidearms, respectively. Standard errors from these repeats are shown. These lifetimes exceed the length of experiments that will take place in these chambers, demonstrating suitable vacuum quality and providing a measure of the pressure at the atoms' location of order $1\times 10^{-10}$~mbar.
    }
    \label{fig:MotLifetime}
\end{figure}

In order to assess the vacuum performance independently from the ion pump values, the loss rate of atoms from a magnetic trap \cite{MagneticTrap} was used to estimate the dominant collision rates within the system, and thus the pressure of the system. Atoms are loaded into a MOT for 4-5 seconds and pumped into the magnetically trappable substates of the \textsuperscript{3}P\textsubscript{2} state by turning off the repump lasers. The atomic beam and 3D MOT light is shut off and the atoms are held in the 3D magnetic trap for a variable hold time, before the repump lasers and 3D MOT light is flashed on. Fluorescence from the remaining atoms is measured with a photodiode and imaging telescope. Figure \ref{fig:MotLifetime} shows the magnetic trap lifetime, where the population decays with a time constant of 28(3)~s and 20(1)~s for the Imperial and Birmingham sidearms, respectively. Using the calibration given in~\cite{MagneticTrap}, the pressure in Chamber 2 is estimated to be $1.0(1) \times 10^{-10}$~mbar and $1.7(1) \times 10^{-10}$~mbar, respectively. The lifetime of the atoms in the magnetic trap far exceeds the length of an experiment in these systems.   

\section{Discussion}
\label{sec:timeline}

The successful delivery of the five UHV systems to the individual institutes in a timely fashion was facilitated by the centralisation of three core activities that were outlined above:
\begin{itemize}
    \item Design;
    \item Manufacture and assembly;
    \item Delivery and logistics.
\end{itemize}
The engineering design process started in April 2020 and ran through to December 2021. During this period the design was taken from a first concept to a fully detailed and analysed design ready for manufacture and commissioning. This centralised design process was led by an engineer at the University of Oxford.  While not necessarily being faster than a traditional design process for an individual setup, crucially the centrally coordinated structure  provided for the required compatibility of the resulting systems that will allow AION to port technologies and cold atom advances between institutions and to the main instruments. It furthermore allowed for several benefits that increased the efficiency of the design engineering process that included:
\begin{itemize}
    \item Resource Optimisation - a centrally-coordinated engineering effort means that the use of engineering resource can be optimised, leading to a reduction in duplication of effort and the ability to ensure that the right people are working on the appropriate tasks, improving productivity and reducing costs;
    \item Standardisation - this increases the speed of the design process, due to a reduction in errors, and the ability to learn from and carry fixes and improvements across systems;
    \item Communication - communications between team members with engineering responsibilities can be improved, reducing and minimising miscommunications. This, in turn, allows a more agile work model and faster decision-making;
    \item Expertise - a common resource pool of engineering talent means that a wider skill set will be available across the project.  This is particularly useful for issues that require specialised skills to resolve, such as UHV design, pumping calculations and opto-mechanics - all of which were present in this phase of the project;
    \item Risk Mitigation - a central point of contact can ensure that better and more consistent controls and mitigation techniques are used to improve risk mitigation.  This, in turn, reduces failures and improves the efficiency of the engineering process.
\end{itemize}
All of the above benefits were realised for the project and improved both the quality of the engineering design work and the rate at which it was completed,
though it is difficult to quantify the time saved during the design phase by the centralised process.

\begin{figure}[t]
    \centering
    \includegraphics[width = 1.0\textwidth]{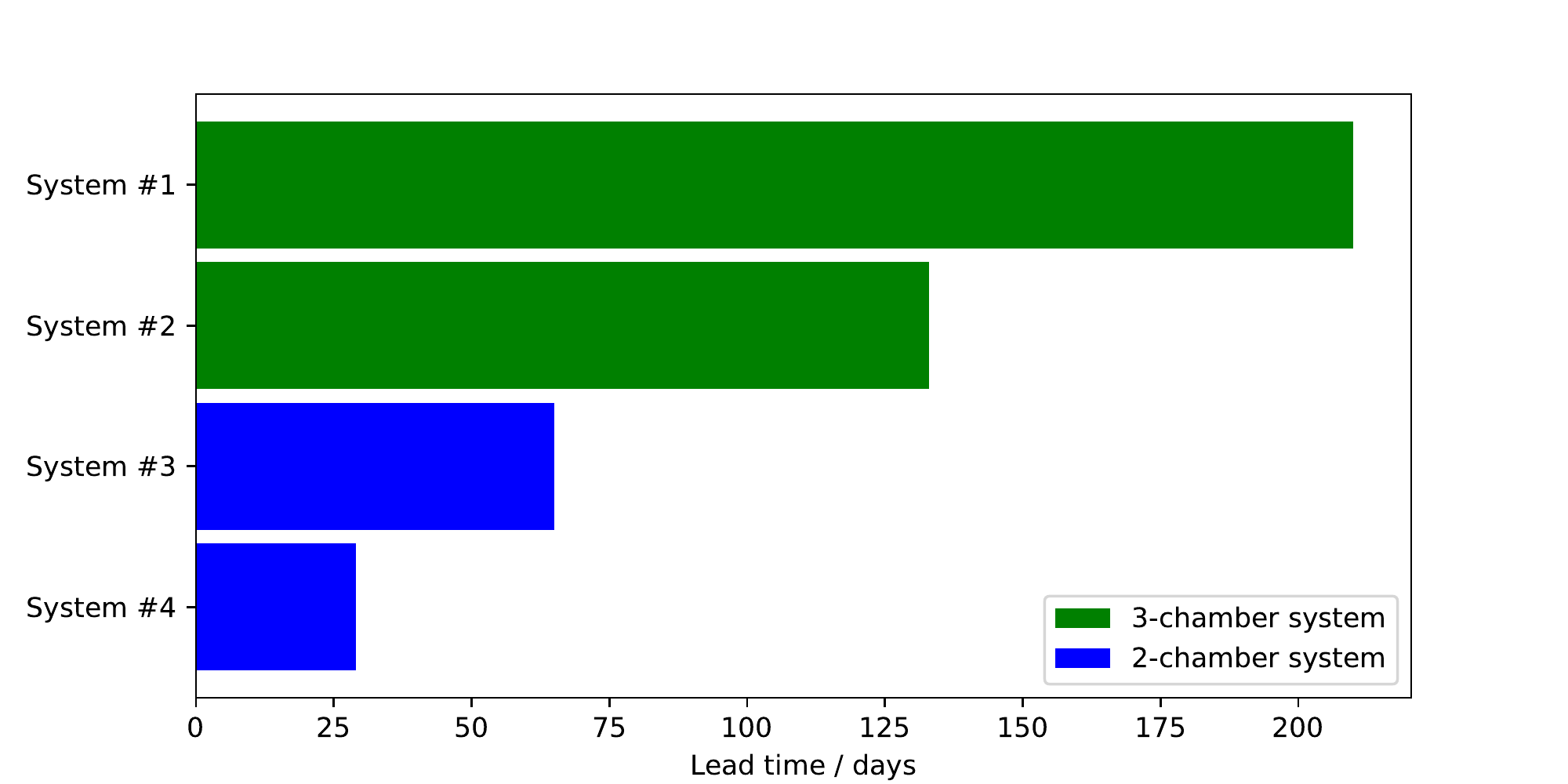}
    \caption{
        Evolution of the build times (in days) of the four centrally-produced UHV systems (3-chamber systems in green, 2-chamber systems in blue) as the project progressed. The build time for
        the Imperial College UHV system is not shown: it was not built centrally since it required specific in-vacuum hardware for quantum squeezing measurements and a specialised build process.}
    \label{fig:Build_Time}
\end{figure}

It is much easier to quantify the efficiency gains for the manufacturing and assembly process.  During this phase of the project, the rate at which a UHV system could be built and commissioned was accelerated greatly.  This was primarily due to carrying the knowledge gained by building one system over into the subsequent builds.  The continuity and accumulation of knowledge would be lost in a non-centralised process. A second, strongly-linked benefit was the creation and development of a facility with a dedicated skillset focused on the requirements of this specific UHV-System build. Figure \ref{fig:Build_Time} shows the build-time duration in days for each system that was built via the centralised process.  The four centralised builds are shown and split into 3-chamber and 2-chamber systems.  The decrease in build-time from system to system can be seen clearly. A more detailed list of reasons why the centralised approach increased the efficiencies of the system builds is as follows:
\begin{itemize}
    \item Continuity of knowledge - the lessons learnt, knowledge and skills gained from the prior builds directly benefited the subsequent builds;
    \item Inventory management - a centrally-managed inventory of long-lead-time parts (such as coated viewports) enabled the utilisation of stock items of later builds in the prior builds when there were failures and difficulties during bake cycles.  The used stock could then be re-ordered in advance, minimising delays on the subsequent systems;
    \item Parallel builds - agile planning and workflow were  facilitated by the centralised, coordinated approach, which enabled parallel system builds to be carried out.  As one system would undergo its bakeout, a second system could be built and assembled.  This yielded large efficiency gains;
\end{itemize}
All of the above contributed to the build-speed increases shown in Fig.~\ref{fig:Build_Time} and made for very significant gains.
The entire engineering process, during which all five UHV systems went from conceptual design to fully delivered systems, took approximately two years and seven months.  Within this period, engineering design took approximately one year and eight months, with manufacture and delivery taking just under a year.

\section{Summary and Conclusion}
\label{sec:summary}

Equipping five ultracold strontium laboratories (UCSLs) was motivated by
the AION plan to grow dedicated cold atom expertise in strontium around the UK and to enable parallel R\&D efforts towards the required design objectives for the
atom sources for later stages of the AION project, including the transport and launch of atoms, Large
Momentum Transfers, squeezing and high-sensitivity atom interferometry. This plan presented the collaboration with
the challenge of building five compatible state-of-the-art UCSLs in parallel,
 which has had no precedent in the field of cold atom physics.

This document has described the approach taken by the AION
collaboration, namely the centralised design and production of
key components of the UCSLs. This approach was familiar to
high-energy physicists, but was a novelty for the cold atom community, where typically every setup has bespoke requirements. It enabled the construction of multiple UCSLs faster and more cost-effectively than following the ``in-house" approach that has typically been taken by the latter community. This document focused on the Ultra-High-Vacuum (UHV), magnet and Laser-Stabilisation (LS) systems, but similar coordinated approaches are also being followed regarding control software and optical design.
Commissioning data on the performances of the UHV, magnet and LS systems obtained by AION UCSL teams have also been presented.

The centralised approach ensured maximal
commonality between the equipment in five state-of-the-art UCSLs
across the UK. Following their construction, the
required R\&D into improved cold-strontium quantum technology can be performed over the next decade.
This approach provided economies of scale in the design and production process,
and is expected to provide benefits also in the
operations of the UCSLs.

The shared approach between the particle and cold atom groups
in the different participating universities and national laboratories has helped
foster a national community combining the expertises of members
of the previously disjoint EPSRC and STFC communities, which
may provide the basis for future valuable cross-fertilisation.

The AION experience could serve as a useful model for the modular and
distributed development and construction of other cold atom
projects, such as atomic clock experiments or neutral atom quantum
computing systems. This model could become increasingly useful as
the range of industrial applications of cold atom systems broadens
in the coming years. International opportunities are already being
explored.
For these reasons, the AION collaboration is investigating the
possibility of commercialising its approach and establishing
dedicated design and/or production units at National Laboratories
such as RAL and Daresbury.

\section*{Acknowledgements}

This work was supported by UKRI through its Quantum Technology for
Fundamental Physics programme, via the following grants from EPSRC and STFC
in the framework of the AION Consortium:
ST/T006536/1 and ST/W006448/1 to the University of Birmingham; ST/T006579/1, ST/W006200/1 and
ST/X004864/1 to the University of Cambridge;
ST/T007001/1 to the University of Liverpool, ST/T006994/1 and ST/W006332/1 to Imperial College London;
ST/T00679X/1 to King's College London; ST/T006633/1 to the University of Oxford;
ST/T006358/1 and ST/W006510/1 to the STFC Rutherford-Appleton Laboratory.
L.B. acknowledges support from the STFC Grant No. ST/T506199/1. 
D.B. is currently supported by a `Ayuda Beatriz Galindo Senior' Grant from the Spanish Ministerio de Universidades, No. BG20/00228. IFAE is partially funded by the CERCA program of the Generalitat de Catalunya.
J.M. acknowledges support from the University of Cambridge Isaac Newton Trust.
J.S. acknowledges support from the Rhodes Trust.
The Cambridge team thanks X.~Su and M.~Zeuner for support in earlier stages of this project.
The Rutherford Appleton Laboratory team thanks S.~Canfer, B.~Green, S.~Greenwood, P.~Jeffery, D.~Tallentire and J.~Tarrant for their valuable input and contributions to the design process of the assemblies and during the building of the sidearms.
For the purpose of open access, the authors have applied for a Creative Commons Attribution (CC-BY) licence to any Author Accepted Manuscript version arising from this submission.

\bibliographystyle{JHEP}
\bibliography{main}

\end{document}